\documentclass[conference]{IEEEtran}
\usepackage{xspace}
\IEEEoverridecommandlockouts
\usepackage{cite}
\usepackage{amsmath,amssymb,amsfonts}
\usepackage{algorithmic}
\usepackage{graphicx}
\usepackage{textcomp}
\usepackage{xcolor}
\usepackage{booktabs}
\usepackage{multirow}
\usepackage{subfiles} % Best loaded last in the preamble
\usepackage{enumitem}
\usepackage{subcaption}
\usepackage{tcolorbox}
\usepackage[hide links]{hyperref}
\usepackage{url}
\usepackage[linesnumbered,ruled,vlined]{algorithm2e}
\usepackage[font=small,labelfont=bf]{caption}

\newcommand{\tool}{IRENE\xspace}

\def\BibTeX{{\rm B\kern-.05em{\sc i\kern-.025em b}\kern-.08em
    T\kern-.1667em\lower.7ex\hbox{E}\kern-.125emX}}
\begin{document}

\title{Integrating Rules and Semantics for LLM-Based C-to-Rust Translation
% From Syntax to Safety: Enhancing LLMs for C-to-Rust Translation via Rule-Based Static Analysis and Structured Summarization
% From Syntax to Safety: Bridging Rule-Based Static Analysis and LLMs for C-to-Rust Translation \\
% Beyond Syntax: Synergize Rules and Semantics for LLM-Based C-to-Rust Translation
% RARS: Rule-Augmented Retrieval and Self-Optimization for C-to-Rust in Industry\\

% \thanks{This research is supported by National Key R\&D Program of China (No. 2022YFB3103900), National Natural Science Foundation of China under project (No. 62472126), Natural Science Foundation of Guangdong Province (Project No. 2023A1515011959), Shenzhen-Hong Kong Jointly Funded Project (Category A, No. SGDX20230116091246007), Shenzhen Basic Research (General Project No. JCYJ20220531095214031), and CCF-Huawei Populus Grove Fund.}
}
% \author{\IEEEauthorblockN{Feng Luo, Kexing Ji, Cuiyun Gao$^*$}
% \IEEEauthorblockA{\textit{School of Computer Science and Technology} \\
% \textit{Harbin Institute of Technology}\\
% Shenzhen, China\\
% hitszluofeng@foxmail.com}
% \and
% \and
% \and
% \and
\author{
Feng Luo$^1$, Kexing Ji$^1$, Cuiyun Gao$^{1*}$, Shuzheng Gao$^2$, Jia Feng$^1$, Kui Liu$^3$, Xin Xia$^4$, Michael R. Lyu$^2$\\
$^1$School of Computer Science and Technology, Harbin Institute of Technology, Shenzhen, China \\
$^2$Department of Computer Science and Engineering, The Chinese University of Hong Kong, China \\
$^3$Huawei Software Engineering Application Technology Lab, China $^4$Zhejiang University, China \\
hitszluofeng@foxmail.com, kexingji@stu.hit.edu.cn, gaocuiyun@hit.edu.cn, szgao23@cse.cuhk.edu.hk, \\ jfeng@std.uestc.edu.cn, kui.liu@huawei.com, xin.xia@acm.org, lyu@cse.cuhk.edu.hk
}

\maketitle

\begin{abstract}
Automated translation of legacy C code into Rust aims to ensure memory safety while reducing the burden of manual migration. 
Early approaches in C-to-Rust translation rely on static rule-based methods, but they suffer from limited coverage due to dependence on predefined rule patterns.
% Early approaches rely on rule-based methods by matching predefined rules, but suffer from limited coverage. 
% and often retain unsafe constructs, such as parts of the original C code. 没必要补充会遗留不安全结构（例如保留了部分原来的C语言）
Recent works regard the task as a sequence-to-sequence problem by leveraging large language models (LLMs). 
% \wenxin{Recent works have introduced large language models (LLMs) to the task, formulating it as a sequence-to-sequence problem.}
Although these LLM-based methods are capable of reducing unsafe code blocks, the translated code often exhibits issues in following
% handling 
Rust rules and maintaining semantic consistency.
% accuracy. 
On one hand, existing methods adopt a direct prompting strategy to translate the C code, which struggles to accommodate the syntactic rules between C and Rust. On the other hand, this strategy makes it difficult for LLMs to accurately capture the semantics of complex code.
% LLMs still lack a deep understanding of the semantic intricacies embedded in source code, making it challenging to produce accurate and executable translations.

% To address these challenges, we propose \tool, an LLM-based framework that synergizes rules and semantics for enhancing LLM translation, consisting of three components: 
To address these challenges, we propose \tool, an LLM-based framework that \textbf{I}ntegrates \textbf{R}ul\textbf{E}s a\textbf{N}d s\textbf{E}mantics to enhance translation. \tool consists of three modules:
%介绍各个模块前，缺少整个框架如何协调使用这三个模块的总结
1) a rule-augmented retrieval module that selects relevant translation examples based on rules generated from a static analyzer developed by us, thereby improving the handling of Rust rules;
2) a structured summarization module that produces a structured summary for guiding LLMs to enhance the semantic understanding of C code;
3) an error-driven translation module that leverages compiler diagnostics to iteratively refine translations.
% , fully exploiting the LLMs' self-correction ability
% Extensive experiments conducted on eight LLMs and two datasets demonstrate the effectiveness of \tool. 
%实验在8个LLM以及2个数据集（1个xcodeeval benchmark，1个huawei内部数据）上开展，主要评估两个方面translation reliability（CA,CSR） and safety（UR,ULR），每个方面又分别对应两个指标。所以实际实验是这么安排的：RQ1：验证xcodeeval上的reliability and safety，其中reliability用5个baseline对比了CA和CSR指标，safety用3个baseline对比了UR和ULR指标； RQ3：reliability and safety都用了5个baseline，对比了CSR，UR和ULR指标，但是由于huawei资源限制，只部署了3个大模型。
We evaluate \tool on two datasets (xCodeEval---a public dataset, HW-Bench---an industrial dataset provided by Huawei)
% private) 
and eight LLMs, focusing on translation accuracy and safety.
In the xCodeEval, \tool consistently outperforms the strongest baseline method in all LLMs, achieving average improvements of 8.06\% and 12.74\% in the computational accuracy (CA) and compilation success rate (CSR), respectively. It also enhances the safety of translated code, reducing the Unsafe Rate (UR) to 1.70\% on average.
In the HW-Bench, when compared to the strongest baseline, \tool improves CSR and reduces UR by an average of 0.33\% and 26.00\%, respectively.

\end{abstract}

% \begin{IEEEkeywords}
% LLM, Rust, Code Translation, Rules.
% \end{IEEEkeywords}

\section{Introduction}

Legacy C projects remain vulnerable to memory safety issues
%continue to struggle with memory safety vulnerabilities 
due to the language’s lack of built-in safety mechanisms~\cite{DBLP:conf/icse/Hong23, DBLP:conf/apsys/ChenMWZZK11, DBLP:conf/imc/DurumericKAHBLWABPP14}. 
A study by Mozilla reveals that 94.00\% high-severity and critical security bugs in Firefox are related to memory safety, with C/C++ being the primary culprits~\cite{herman2019rewriting}.
Rust~\cite{rustforlinux_nova} offers memory safety without garbage collection, combining the low-level control of C/C++ with the ease of use of modern languages. 
% Consequently, it is being adopted by major companies to enhance the safety and reliability of system-level software.
For example, Nvidia has reimplemented the GPU driver Nova in Rust, and Google has adopted Rust for new Android operating system components~\cite{google2021rust}.
%Notable examples include Nova, an Nvidia GPU driver reimplemented in Rust~\cite{rustforlinux_nova}, and Google's adoption of Rust in Android for developing new operating system components~\cite{google2021rust}. 
However, manual C-to-Rust migration presents significant challenges, including Rust's steep learning curve and the critical requirement to preserve the original program's semantics. 
%rewriting C code in Rust is challenging due to the steep learning curve and the need to preserve program semantics. 
Consequently, the task of C-to-Rust translation, which aims at automatically generating semantically equivalent Rust code, has emerged as a promising direction among academia and industry.
% for research.

% Currently, these methods can be broadly categorized into two types: static rule-based methods and learning-based methods. Static rule-based methods[] directly apply rules for translation. 
Early work on C-to-Rust translation has mainly centered around rule-based approaches that rely on predefined patterns to guide the translation process~\cite{DBLP:conf/cav/ZhangDYW23, DBLP:conf/kbse/HongR24, DBLP:conf/icse/LingYWWCH22}. A representative example is C2Rust~\cite{immunant2022c2rust}, a static tool built on Clang~\cite{clang} and LLVM~\cite{DBLP:conf/cgo/LattnerA04} that systematically converts C code into Rust. Subsequent methods extend C2Rust to better support specific translation targets,
% handle target types,\yun{[unclear??]}
such as parallel APIs~\cite{concrat} and pointer constructs~\cite{ownership,Emre2021,Emre2023}.
These rule-based methods offer the advantage of high efficiency and low computational cost. However, they heavily rely on predefined rules created by experts, which are labor-intensive and suffer from the low coverage problem~\cite{Emre2021,Hong_static}.
% Consequently, methods that apply LLMs to this task have emerged.
% large language models (LLMs) have achieved promising results in code-related tasks, including code translation[]. 

%这一段主要是介绍LLM-based方法并介绍motivation，LLM-based方法面临的困难就是我们为什么要这样做的动机，主要是两个困难，一个是对规则follow的不准确，我展示了两个case；另一个是对语义理解不足，我则是引用了其他文章的结论。
\begin{figure*}
    \centering
    \includegraphics[width=0.85\textwidth]{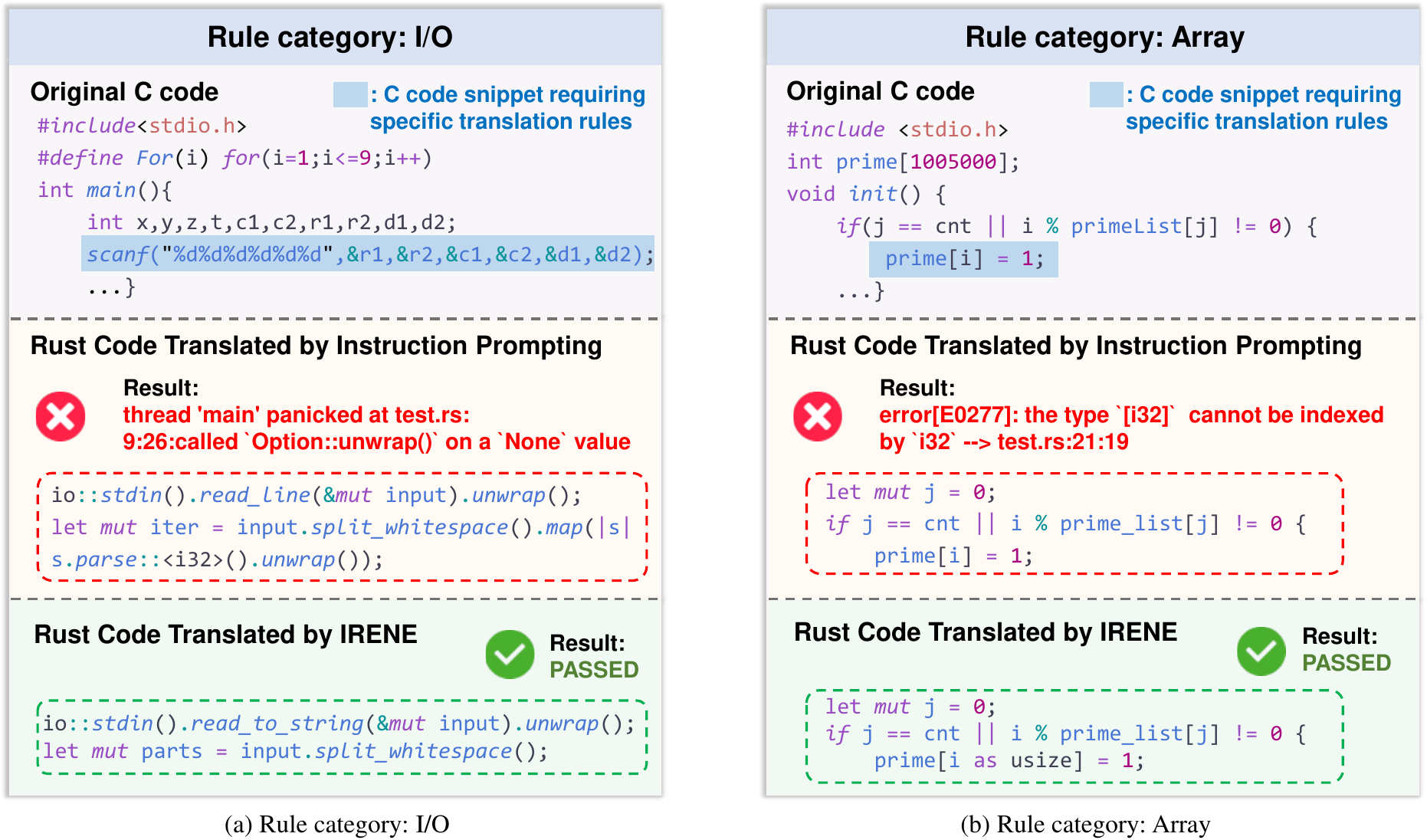}
    \caption{Two cases show that the Rust code translated by instruction prompting fails to handle I/O operations (a) and array types (b) properly based on Qwen2.5-Coder-32B.
    % how \tool improves translation quality by following Rust-specific rules.
% (a) In the I/O case, instruction prompting fails to handle multiple variable inputs correctly, resulting in a runtime panic. \tool adopts a rule-guided strategy using read\_to\_string and split\_whitespace, which ensures semantic equivalence and avoids the error.
% (b) In the Array case, the baseline produces a type mismatch error by indexing with i32. \tool resolves this by converting the index to usize, conforming to Rust's array indexing rule.
}
    \label{rule_1}
    \vspace{-1em}
\end{figure*}
%In contrast, 
Recently, LLM-based methods have emerged as a promising paradigm for code translation~\cite{DBLP:journals/corr/abs-2409-10506, DBLP:journals/corr/abs-2412-14234, DBLP:journals/corr/abs-2501-14257}. These approaches learn translation patterns from large-scale code corpora and aiming at generating
% generate 
safe Rust code compared to rule-based approaches~\cite{VERT,crownllm}. %are capable of learning translation patterns from large-scale code corpora and have shown the ability to produce safer Rust code. 
% However, these methods solely rely on the inherent capability of LLMs without incorporating explicit guidance for code safety, which often results in code that violates Rust's rules.
%rely on direct prompting, which often leads to issues in following Rust rules. 
However, these methods solely rely on the inherent capability of LLMs without explicit guidance, which often results in code that violates Rust's rules.
For example, as illustrated in Fig.\ref{rule_1}, the Rust code translated by instruction prompting (i.e., directly asking the model to translate with a general instruction) fails to handle I/O operations and array types properly, violating Rust’s safety requirements.
Some recent studies\cite{C2S,Multi} have attempted to integrate static rule-based analysis to improve LLMs' performance. 
% However, these approaches typically employ rules in a superficial manner without providing concrete translation examples
% \yun{[low novelty??]}
However, these approaches frequently employ static tools like C2Rust rather than establishing verified safe translation paradigms, which compromises the quality of translated code and yields insufficient safety guarantees.
% suboptimal or incorrect code generation in complex scenarios.
% as illustrated in Fig.~\ref{rule_1} and Fig.~\ref{rule_2}. A more detailed discussion can be found in Section \textit{Discussion}.
Moreover, LLMs often struggle to comprehensively understand the semantics of complex source programs, particularly when explicit descriptions of I/O types are absent. %Second, according
According to Yang et al.~\cite{yang2024}, more than 50.00\% of translation failures by large language models stem from insufficient understanding of the source code and unclear specification of I/O types. These limitations greatly
% \yun{greatly}
% significantly 
hinders the practical deployment of LLMs in the task.
% \yun{the task.}
% automated code translation tasks.
% are attributed to two main causes: 38.51\% stem from insufficient understanding of the source code, and 14.94\% from unclear specification of I/O types. This highlights that LLMs often struggle to deeply understand the semantics of complex source programs, particularly in the absence of an explicit description of I/O types. 
% These limitations can lead to incorrect translations, undermining the reliability of such methods in practical scenarios.

To address these challenges, we propose \tool, a rule-augmented retrieval and structured summarization LLM-based framework for C-to-Rust translation. \tool synergistically integrates rules and semantics,
% Specifically, \tool leverages static analysis rules to guide the retrieval of semantically relevant code examples, which serve as informative prompts to enhance LLM translation. This combination allows the framework to enforce domain-specific translation constraints while benefiting from the generalization capabilities of large models. 
consisting of three key components:

\begin{itemize}
\item \textbf{Rule-Augmented Retrieval Module}: This module retrieves relevant translation pairs based on a set of static rules that reflect Rust-specific memory and type semantics, such as input handling conventions, and array indexing patterns. These pairs provide targeted context that helps the model adhere to Rust's strict rules.

\item \textbf{Structured Summarization Module}: 
%分支策略并没有很好的体现
% \wenxin{Inspired by the divide-and-conquer strategy,} 
This module prompts the model to first generate a summary of the C code that describes its I/O types and overall functionality, aiming to produce more semantically faithful translations.

\item \textbf{Error-Driven Translation Module}: After the initial translation, this module parses compiler diagnostic messages and uses them to guide iterative refinement
% regeneration 
of the Rust code. This feedback cycle
% loop
leverages the LLM's ability to learn from error messages and improve its output.
\end{itemize}

To validate the performance of \tool, we conduct experiments on 
a public and an industrial dataset (xCodeEval~\cite{xcode} and HW-Bench). We compare \tool with five representative prompt baselines on eight popular open-source LLMs. Extensive experimental results demonstrate the effectiveness of \tool. \tool delivers consistent and substantial improvements across both datasets.
On the xCodeEval benchmark, we observe that \tool outperforms the strongest baseline across all LLMs. It yields average improvements of 8.06\% in computational accuracy (CA) and 12.74\% in compilation success rate (CSR). Moreover, \tool contributes to safer translations by notably reducing the Unsafe Rate (UR) and Unsafe Loc Rate (ULR) to 1.70\% and 0.26\% on average, outperforming both static analysis tools and prompting methods.
Our evaluation of \tool on the Huawei-developed industrial dataset HW-Bench exhibits consistent performance trends. When compared to the strongest baseline, \tool raises CSR by 0.33\%, and reduces UR and ULR by 26.00\% and 10.49\% on average, respectively.
% , across all employed LLMs.
% On the public dataset xCodeEval~\cite{xcode}, it reliably outperforms all baseline methods, achieving average gains of 10.84\% in the success-translated metric and 56.35\% in the safety metric across multiple models. On the industrial dataset, \tool also yields significant benefits, with an average xx\% increase in compilation success rate and a xx\% reduction in the rate of unsafe code. 

% Our results show that each module contributes significantly to the overall performance, and that the full framework produces code that is both safer and more correct than baselines. Notably, \tool achieves a lower unsafe usage rate while maintaining high translation fidelity.

In summary, our main contributions are as follows:
\begin{itemize}
\item %We identify key challenges in LLM-based C-to-Rust translation, focusing on safety-aware rule adaptation and semantic understanding.
We identify two essential challenges in C-to-Rust translation: the need for safety-aware rule adaptation and accurate semantic understanding
\item 
% 这个贡献讲得不够清晰integrates retrieval, summarization, and refinement讲得云里雾里，太突兀了
%We introduce \tool, a feedback-driven translation framework guided by rules and semantics, which integrates retrieval, summarization, and refinement.
We propose \tool, a novel LLM-based framework for C-to-Rust translation that integrates rule-augmented retrieval, structured summarization, and iterative refinement. Guided by safety-aware translation rules and semantic understanding, \tool systematically improves translation quality and ensures compliance with Rust’s strict safety guarantees.
\item Extensive experiments are conducted on multiple LLMs and datasets, showing that \tool notably outperforms all baselines in both translation accuracy
% \yun{accuracy}
% reliability 
and safety.
\end{itemize}

% \section{Related Work}
% \input{Sections/2_related_work}

\section{Methodology}
\begin{figure*}
    \centering
    \includegraphics[width=1\textwidth]{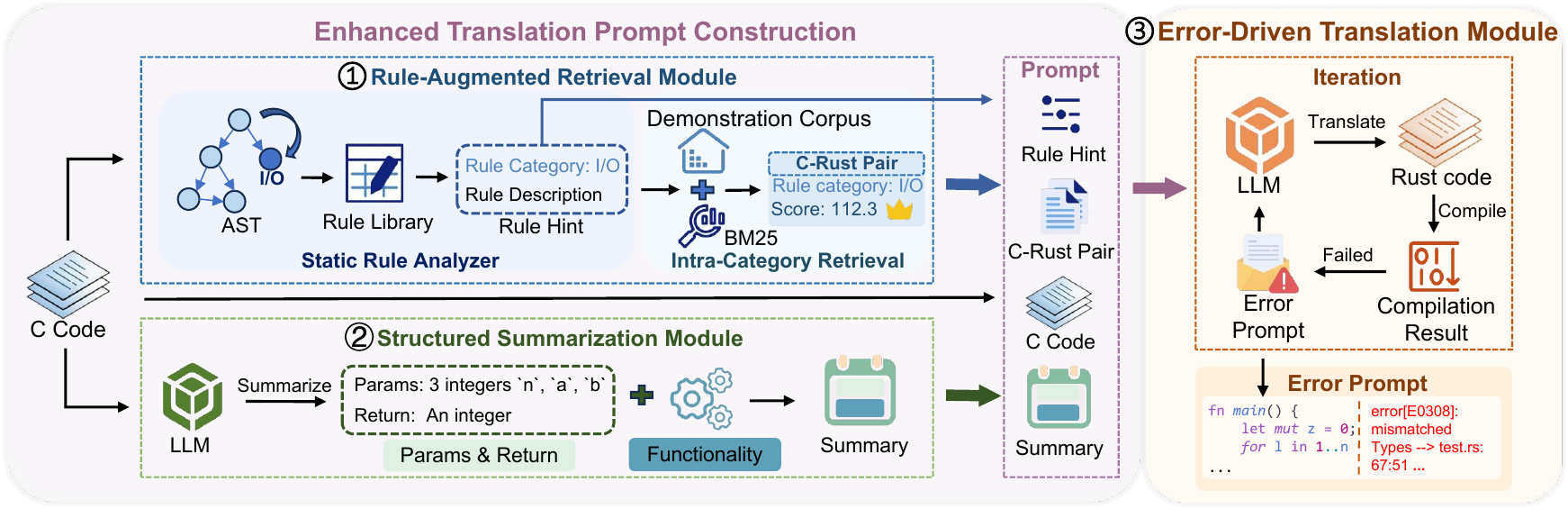}
    \caption{Overview of \tool. }
    \label{architecture}
    \vspace{-1em}
\end{figure*}

\begin{figure}
    \centering
    \includegraphics[width=0.5\textwidth]{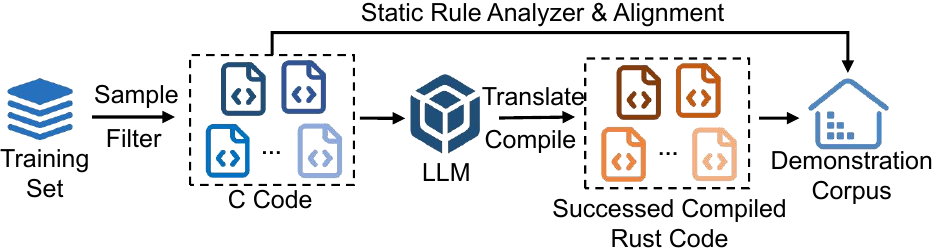}
    \caption{Demonstration corpus construction.}
    \label{corpus}
    \vspace{-1em}
\end{figure}

% In this section, we describe the overall framework of the proposed \tool. As illustrated in Fig.~\ref{architecture}, \tool consists of three main components as detailed below.

In this section, we propose \tool, an LLM-based framework that integrates rules and semantics to enhance translation.
We first present the overview of \tool and then elaborate on its module details in the following subsections.
%  As shown in Fig. 2, the framework is composed of three main components, each of which is detailed below.
%这里要说明三个模块间的协作关系
\subsection{Overview}
% To address the main challenges of C-to-Rust translation, we design a pipeline composed of three modules.
As illustrated in Fig.~\ref{architecture}, \tool consists of three main components organized into a pipeline.
First, the \textbf{rule-augmented retrieval module} selects relevant translation examples $E_{demo}$ based on syntactic rules
%\(\text{Rule}_{\text{hint}}\) 
% (\\text{R}_{hint}\)
$R_{hint}$
extracted from our static analyzer, aiming at reinforcing Rust's safety rules. Next, the \textbf{structured summarization module} generates a structured semantic summary \(S_c\) of the C code to improve the LLMs’ understanding and translation quality. 
%These outputs—\(E_{\text{demo}}\), \(\text{Rule}_{\text{hint}}\), \(S_c\), along with the original C code \(\text{CODE}_c\)—are composed into a prompt that guides the LLM to produce initial Rust translations. 
These outputs (\(E_{\text{demo}}\), \(R_{\text{hint}}\), \(S_c\)), along with the original C code (\(CODE_c\)) are composed into a prompt that guides the LLM to produce initial Rust translations.
Finally, the \textbf{error-driven translation module} utilizes compiler diagnostics to iteratively refine the output, correcting translation errors and enhancing safety. 
\begin{algorithm}[h]
\small
\caption{Static Rule Analyzer---DetectPointer}
\label{detect_method}
\KwIn{C source code $CODE_{c}$}
\KwOut{Set of $R_{hint}$ triplets}

\Begin{
    Parse $CODE_{c}$ into AST tree\;
    \SetKwProg{Fn}{Function}{:}{}
    \Fn{Walk(node)}{
        \If{node is pointer related}{
            decl\_node $\leftarrow$ Find the nearest enclosing declaration\;
            rule\_category $\leftarrow$ "Pointers"\;
            code\_snippet $\leftarrow$ extract corresponding code from $CODE_{c}$\;
            suggested\_Rust $\leftarrow$ \textsc{InferPointerType}(code\_snippet)\;
            rule\_description $\leftarrow$ code\_snippet+suggested\_Rust\;
            Append \{rule\_category, rule\_description\} to $R_{hint}$\;
        }
        \ForEach{child in node.children}{
            \textsc{Walk}(child)\;
        }
    }

    \textsc{Walk}(\texttt{AST.root})\;
    \Return $R_{hint}$\;
}
\end{algorithm}
\subsection{Rule-Augmented Retrieval Module}
This module
% \yun{module}
% component 
is designed to help LLMs adapt to Rust’s strict safety rules.
% \wenxin{This component plays a foundational role by supplying the LLM with detailed rule hint, enabling it to internalize and adhere to Rust’s strict safety principles during translation.}
% \kexing{This component is intended to support the alignment of large language models with Rust’s strict safety requirements.}
First, we develop a static rule analyzer tailored to Rust’s memory and type safety requirements, which extracts necessary translation rules from the input 
%C code.
\(CODE_c\).
Then, based on the identified rules, we retrieve the most relevant translation examples from a curated demonstration repository. These examples help the model understand how abstract rules are concretely applied during translation. In this way, the rules serve as a high-level guideline, while the examples provide practical demonstrations, enabling the model to learn rule-to-code mappings more effectively.
\subsubsection{Static Rule Analyzer}
Building upon previous studies~\cite{repo,Emre2021},
% and our empirical observations of LLMs applied to C-to-Rust translation \yun{[???]}, 
we summarize and extend a translation rule library to enhance automation and consistency. A representative subset of the rule library is presented in Table~\ref{rules}, serving as guidance for the transformation process. Then, we design a corresponding rule detection algorithm. The detection method for pointer-related rules is illustrated in Algorithm~\ref{detect_method}, and similar strategies are applied to other rule categories. The detection algorithm leverages the abstract syntax tree (AST) and outputs the rule hint $R_{hint}$, each represented as a triplet: 
\begin{equation}
    R_{hint} = ⟨rule\ category, rule\ description⟩
\end{equation}
where the $rule\ category$ indicates the type of transformation rule detected, the $code\ snippet$ refers to the minimal C fragment that requires the application of the rule, and the $suggested\ Rust$ is the Rust code recommended by the rule. Since $R_{hint}$ is concise, it serves as an initial hint and requires further augmentation via few-shot retrieval in the next step.

% The algorithm has a time complexity of $O(n)$, where $n$ is the number of nodes in the AST, ensuring scalability for large codebases. 

% \usepackage{booktabs}
% \usepackage{multirow}
% \begin{table*}[]
% \centering
% \caption{This table presents the taxonomy of rule categories designed in our Static Rule Analyzer. Due to space limitations, we show only a subset of the rule descriptions and corresponding examples. Each \textbf{rule description} summarizes the recommended translation strategy, while the \textbf{example} column illustrates the actual translation of a representative C code snippet to Rust.
% When multiple rules are shown in the table, we annotate the corresponding examples with matching indices to indicate their alignment.}
% \label{rules}
% % \includegraphics[width=0.8\linewidth]{Figures/case_IO.pdf}
% \end{table*}
\begin{table*}[]
\caption{This table presents the taxonomy of rule categories designed in our Static Rule Analyzer. We present one representative rule for each category. The complete rules can be found in our GitHub repository~\cite{IRENE}. Each \textbf{rule description} summarizes the recommended translation strategy, while the \textbf{example} column illustrates the actual translation of a representative C code snippet to Rust.}
\label{rules}
\resizebox{\linewidth}{!}{%
\begin{tabular}{@{}c|l|ll@{}}
\toprule
\multirow{2}{*}{\textbf{Category}} & \multicolumn{1}{c|}{\multirow{2}{*}{\textbf{Description}}} & \multicolumn{2}{c}{\textbf{Examples}} \\ \cmidrule(l){3-4} 
 & \multicolumn{1}{c|}{} & \multicolumn{1}{c|}{\textbf{C Code Snippets}} & \multicolumn{1}{c}{\textbf{Rust Code Snippets}} \\ \midrule
\textbf{Pointers} & Use Box\textless{}T\textgreater for malloc-allocated structs. & \multicolumn{1}{l|}{struct Node* n = malloc(sizeof(struct Node));} & \begin{tabular}[c]{@{}l@{}}struct Node \{ /* fields */ \};\\ let n = Box::new(Node::default());\end{tabular} \\ \midrule
\textbf{I/O} & Replace scanf with read\_to\_string() + split\_whitespace() & \multicolumn{1}{l|}{\begin{tabular}[c]{@{}l@{}}int a; float b;\\ scanf("\%d \%f", \&a, \&b);\end{tabular}} & \begin{tabular}[c]{@{}l@{}}let mut input = String::new();\\ io::stdin().read\_to\_string(\&mut input).unwrap();\\ let mut parts = input.split\_whitespace();\end{tabular} \\ \midrule
\textbf{Mixtype} & Cast mixed-width integers to a common type like i64 & \multicolumn{1}{l|}{\begin{tabular}[c]{@{}l@{}}int sum = 0;\\ sum = sum + (long long) dp{[}i{]} * dq{[}idx{]} ;\end{tabular}} & \begin{tabular}[c]{@{}l@{}}let mut sum: i64 = 0;\\ sum = sum + (dp{[}i{]} as i64 * dq{[}idx{]} as i64);\end{tabular} \\ \midrule
\textbf{Array} & \begin{tabular}[c]{@{}l@{}}Cast array indices to usize in Rust and\\ use {[}{[}T; M{]}; N{]} for fixed-size multidimensional arrays.\end{tabular} & \multicolumn{1}{l|}{\begin{tabular}[c]{@{}l@{}}int dp{[}100{]}{[}2{]};\\ res = res + dp{[}i + l{]}{[}(l - 1) \& 1{]};\end{tabular}} & \begin{tabular}[c]{@{}l@{}}let dp = {[}{[}0; 2{]}; 100{]};\\ res += dp{[}(i + l) as usize{]}{[}((l - 1) \& 1) as usize{]};\end{tabular} \\ \bottomrule
\end{tabular}
}
\end{table*}

\subsubsection{Intra-Category Retrieval}
This step aims to provide further augmentation for the rule-based translation process. While the detection stage yields a suggested Rust type as part of $R_{hint}$ this suggestion is often partial or abstract. To bridge this gap, we retrieve concrete, structurally similar translation examples from the demonstration corpus, offering richer contextual information to guide the model's translation.

We first construct a demonstration corpus to support retrieval-based translation. As shown in Fig.~\ref{corpus}, the construction process involves random sampling and a two-step filtering pipeline. Specifically, we randomly sample 30,000 instances from the xCodeEval~\cite{xcode} training set. To prevent data leakage, we apply an initial filter. We then use the Qwen2.5-Coder-32B model to generate the corresponding Rust code and apply a second filter by compiling the generated Rust code to ensure successful compilation. After filtering, we aligned the C and Rust code to obtain 22,892 translation pairs in the final corpus. We further annotate each sample in the corpus with rule categories using our Static Rule Analyzer.
For retrieval, we adopt BM25~\cite{BM25}, a classic sparse retrieval algorithm from the information retrieval domain, known for its effectiveness in capturing syntactic similarity between code snippets. Since our approach emphasizes rule-level similarity, BM25 provides an effective mechanism for identifying structurally similar examples.

To support translation via demonstration, we retrieve examples $E_{demo}$ from the corpus that match the rule category of the input $CODE_{c}$. Each retrieved example is denoted as
\begin{equation}
    E_{demo}=⟨c\ code,translated\ rust\ code⟩
\end{equation}

\subsection{Structured Summarization Module}
% 这个模块的本意是为了增进模型对源代码语义的理解和认知，确保翻译得到的代码能保持原有的功能。
We introduce a structured summarization module that prompts the model to first produce a structured summary of the C code, thus ensuring that the subsequent Rust translation remains semantically faithful.
% \wenxin{The Structured Summarization Module helps bridge the semantic gap by guiding the model to first produce a structured summary of the C code, thus ensuring that the subsequent Rust translation remains semantically faithful.}
The key motivation is to extract semantically relevant structural and type information from the $CODE_{c}$, in a form that is concise, interpretable, and beneficial for downstream rule-based translation. This module aims to transform low-level C code into a higher-level description that emphasizes semantic roles. The output acts as an intermediate representation for guiding example retrieval and supporting rule-aware translation.
We denote this summarization process as a triplet:
\begin{equation}
    S_{c}=⟨Input,Output,Functionality⟩
\end{equation}

Based on the outputs of the previous modules, we construct an enhanced translation prompt that integrates multiple signals:
\begin{equation}
    prompt = ⟨R_{hint}, E_{demo}, CODE_{c}, S_{c}⟩
\end{equation}
This unified prompt provides the model with rule-level intent, retrieval-based demonstrations, and a high-level summary of the input code. Together, these elements guide the LLM towards
% toward 
producing more accurate, idiomatic, and semantically aligned Rust code.
\subsection{Error-Driven Translation Module}
%This module aims to fully leverage the LLM’s capability for self-correction, guided by feedback from the Rust compiler. 
This module is designed to exploit the LLM’s capacity for self-correction, using feedback from the Rust compiler to iteratively refine the generated Rust code.
After the initial Rust translation is generated, we first compile it using the rustc compiler. If any compilation errors are encountered, the error messages are extracted and fed back into the model along with the erroneous Rust output.
The LLM is then prompted to revise the translation, explicitly addressing the compiler diagnostics. This targeted approach enables the model to focus on specific issues, such as type mismatches, borrow checker violations, or missing imports, without regenerating the entire code.

This module acts as a feedback cycle between the compiler and the LLM. By incorporating concrete error diagnostics, it enhances the accuracy and safety of the final translation with minimal human intervention. 
% In practice, we find that one refinement round is typically sufficient to produce more compilable and idiomatic Rust code.

\section{Experimental Setup}

\subsection{Research Questions}
% In order to evaluate \tool, we answer the following research questions:
To evaluate \tool's performance in C-to-Rust translation, we explore the following research questions:
\begin{enumerate}[label=\bfseries RQ\arabic*:,leftmargin=.5in]
    % \item How effective and safe is \tool in translating C to Rust code?
    \item How does \tool perform in terms of effectiveness and safety in C-to-Rust translation?
    
    % We evaluate both translation accuracy and safety improvements.
    \item What is the impact of each component of \tool on the overall translation performance?
    
    % We conduct ablation studies to measure the contribution of retrieval, summarization, and refinement.
    \item How does \tool perform in industrial scenarios?

    % We test \tool on industry-grade C projects.

\end{enumerate}

\subsection{Datasets}

\textbf{Few-Shot Retrieval Corpus.}
We build a retrieval corpus composed of 22892 C-Rust translation pairs. Each example includes both the original C code and the corresponding Rust implementation. The corpus is used to provide few-shot demonstrations for LLMs inference. 

\textbf{Evaluation Set.}
We evaluate \tool on two datasets: xCodeEval and HW-Bench.
For xCodeEval, we select 515 samples from the evaluation set, where each C code sample passes all test cases and has a corresponding Rust implementation. 
% These C and Rust versions solve the same problems but in different programming languages, enabling a fair assessment of translation correctness.
For the HW-Bench, we randomly sample 100 C functions from proprietary projects provided by Huawei, covering a diverse set of industry scenarios.

\subsection{Baseline Methods}
In C-to-Rust translation experiments, we compare \tool with baselines as detailed below. \textbf{Instruction} provides the C code along with the prompt ‘Translate the following C code to Rust’ without any additional demonstration. \textbf{In-context Learning (ICL)~\cite{DBLP:conf/nips/BrownMRSKDNSSAA20}} includes several translation examples in the prompt to help the model generate the target Rust code. \textbf{Retrieval-augment generation (RAG)~\cite{DBLP:conf/kdd/FanDNWLYCL24}} retrieves examples from the corpus similar to the target translation c code and includes them in the input prompt. \textbf{Chain-of-though (CoT)~\cite{DBLP:conf/nips/Wei0SBIXCLZ22}} prompts the LLM to explain the translation steps before generating the Rust code, with demonstration examples identical to those in ICL. \textbf{Vert~\cite{VERT}} uses an LLM with few-shot examples and a two-stage refinement process. It first applies compiler-suggested fixes to the source code, then feeds the updated code and diagnostics back into the LLM for further correction. \textbf{C2Rust~\cite{immunant2022c2rust}} is a widely used static rule tool that translates C code into Rust by using predefined rules.

\subsection{Implementation Details}

% 这里逻辑有点混乱，我觉得这样写就行了：
% 1.我们评估RARS用了哪几个模型，利用Pytorch和HF怎么实现的
% 2.  对于Instruction，xxx
%     对于ICL，xxx
%     对于RAG，xxx
%     ...
% 3.实验参数，temperature，top-p, demo number。
% 4.所有实验在哪开展的

\textbf{LLM selection and setup.}
We evaluate \tool on xCodeEval with eight open-source LLMs, DeepSeek-Coder~\cite{Deepseekcoder}, Qwen2.5-Coder-Instruct~\cite{hui2024qwen2}, and Codellama~\cite{Codallama}, in response to the first two RQs. For RQ3, due to resource and time constraints, we limit our evaluation on HW-bench to the Qwen2.5 series models only. We download these models from HuggingFace~\cite{huggingface} and utilize PyTorch~\cite{pytorch} for deploying them locally and performing inference.

\textbf{The implementation of baselines and \tool.}
For ICL, we follow the prior work~\cite{DBLP:conf/kbse/GaoWGWZL23} and randomly sample four examples from the corpus.
For RAG, we retrieve the most relevant example using the BM25 algorithm.
For Vert, we directly use the replication packages released by the authors and adapt them to our tasks. 
For C2Rust, we use the C2Rust 0.20.0 version.
For \tool, we select the example with the highest similarity and apply one iteration of refinement. Following prior work~\cite{threshold1}, we set a BM25 threshold of 100 based on top-1 score distribution, covering 52.04\% of the queries to ensure high-quality examples.

\textbf{Hyperparameter and environment settings.}
As for the hyperparameters of generation, we configure all LLMs with greedy decoding (temperature = 0, top-p = 1) to produce deterministic outputs. We execute Python programs with Python 3.8.20 and compile all Rust programs with Rust 1.81.0. All experiments are conducted on an Ubuntu-20.04 server equipped with four NVIDIA A100 GPUs. 

% As for the hyperparameters for the generation of \LCM{s}, we configure with greedy decoding (temperature 0, top-p 1) to yield deterministic output.

% All experiments are conducted on an Ubuntu-20.04 server equipped with four NVIDIA A100 GPUs.
% % The system runs Ubuntu 20.04 with Linux kernel version 5.15.0-102-generic. 
% We utilize PyTorch and Hugging Face Transformers for deploying LLMs locally and performing inference.
% % Our implementation leverages PyTorch and Hugging Face Transformers for model inference.

% We evaluate \tool on eight popular open-source large language models, including DeepSeek-Coder (1.3B and 6.7B), Qwen2.5-Coder-Instruct (1.5B, 7B, 14B, and 32B), and StarCoder2 (7B, and 15B). 

% All models are configured with greedy decoding (temperature 0, top-p 1) to yield deterministic output.

% Each prompt is evaluated independently across models using a single few-shot example. Compiler-driven refinement is applied with a maximum of one iteration to simulate practical usage under minimal overhead.

\subsection{Performance Metrics}

% We use the following four widely used performance metrics in our evaluation:
To evaluate the translation accuracy, we use the following two widely used performance metrics~\cite{yang2024} in our evaluation:

\textbf{Computational Accuracy (CA)} 
evaluates whether the candidate translation generates the same outputs as the reference when given the same inputs. It can be formally defined as:
\begin{align}
\text{CA} &= \frac{\sum_{k=1}^{N} ca(y_k, \hat{y}_k)}{N} \\
ca(y_k, \hat{y}_k) &=
\begin{cases}
1, & \text{if } Exec_k(y_k) = Exec_k(\hat{y}_k) \\
0, & \text{otherwise}
\end{cases}
\end{align}
where $N$ is the total number of evaluated code samples. $y_k$ and $\hat{y}_k$ represent the ground truth and the generated translation for the $k$-th sample, respectively. $E_{x \in C_k}(\cdot)$ denotes the execution result of a program on the $k$-th test case. The indicator function $\mathrm{ca}(y_k, \hat{y}_k)$ equals 1 if both outputs match, and 0 otherwise.

\textbf{Compilation Success Rate (CSR)}
measures the percentage of translated programs that successfully compile without errors.

To evaluate the translation safety, following prior work~\cite{Multi}, we use the metrics below:

\textbf{Unsafe Rate (UR)} refers to the proportion of unsafe samples in the dataset. 
% It can be formally defined as:
% \[
% {Unsafe\_Rate} = \frac{N_{\text{unsafe\_samples}}}{N_{\text{total\_samples}}}
% \]
% where $N_{\text{unsafe\_samples}}$ denotes the number of samples containing at least one \emph{unsafe} block, $N_{\text{total\_samples}}$ is the total number of samples.

\textbf{Unsafe Loc Rate (ULR)} denotes the ratio of unsafe lines of code within each individual sample. 
% It can be formally defined as:
% \[
% {Unsafe\_Loc\_Rate} = \frac{L_{\text{unsafe}}}{L_{\text{total}}}
% \]
% where $L_{\text{unsafe}}$ represents the number of lines marked as \emph{unsafe} within a sample, and $L_{\text{total}}$ is the total number of lines in that sample.

\section{Experimental Result}

\begin{table*}[]
\centering
\vspace{-1em}
\caption{Comparison between \tool and baselines on computational accuracy (CA) and compilation success rate (CSR). Under each metric, the best performance is highlighted in bold.}
\label{effectiveness}
\begin{tabular}{@{}l|cccc|cccccccc|cccc@{}}
\toprule
     & \multicolumn{4}{c|}{Deepseek-Coder} & \multicolumn{8}{c|}{Qwen2.5-Coder} & \multicolumn{4}{c}{Codellama} \\
 \cmidrule(lr){2-5} \cmidrule(lr){6-13}  \cmidrule(lr){14-17}
\multicolumn{1}{c|}{\textbf{Approach}} & \multicolumn{2}{c|}{1.3B} & \multicolumn{2}{c|}{6.7B} & \multicolumn{2}{c|}{1.5B} & \multicolumn{2}{c|}{7B} & \multicolumn{2}{c|}{14B} & \multicolumn{2}{c|}{32B} & \multicolumn{2}{c|}{7B} & \multicolumn{2}{c}{34B} \\
& CA & \multicolumn{1}{c|}{CSR} & CA & CSR & CA & \multicolumn{1}{c|}{CSR} & CA & \multicolumn{1}{c|}{CSR} & CA & \multicolumn{1}{c|}{CSR} & CA & CSR & CA & \multicolumn{1}{c|}{CSR} & CA & CSR \\ \midrule
Instruction & 4.85 & \multicolumn{1}{c|}{20.78} & 19.03 & 43.69 & 6.99 & \multicolumn{1}{c|}{15.15} & 38.06 & \multicolumn{1}{c|}{67.57} & 45.44 & \multicolumn{1}{c|}{74.17} & 59.03 & 87.18 & 2.91 & \multicolumn{1}{c|}{7.57} & 2.14 & 7.57 \\

ICL & 6.21 & \multicolumn{1}{c|}{64.08} & 27.96 & \multicolumn{1}{c|}{58.83} & 9.51 & \multicolumn{1}{c|}{29.32} & 34.95 & \multicolumn{1}{c|}{63.50} & 48.35 & \multicolumn{1}{c|}{75.53} & 57.86 & \multicolumn{1}{c|}{82.33} & 13.40 & \multicolumn{1}{c|}{41.75} & 15.92 & 39.81 \\

RAG & 6.41 & \multicolumn{1}{c|}{\textbf{70.68}} & 30.49 & \multicolumn{1}{c|}{63.88} & 9.32 & \multicolumn{1}{c|}{41.36} & 34.56 & \multicolumn{1}{c|}{63.69} & 53.20 & \multicolumn{1}{c|}{79.03} & 61.17 & \multicolumn{1}{c|}{84.27} & 14.17 & \multicolumn{1}{c|}{39.22} & 16.12 & 40.78 \\

COT & 6.41 & \multicolumn{1}{c|}{64.85} & 28.35 & \multicolumn{1}{c|}{57.67} & 9.90 & \multicolumn{1}{c|}{33.98} & 34.17 & \multicolumn{1}{c|}{64.66} & 52.23 & \multicolumn{1}{c|}{79.42} & 58.64 & \multicolumn{1}{c|}{85.44} & 12.82 & \multicolumn{1}{c|}{41.17} & 17.09 & 43.50 \\

Vert & 1.55 & \multicolumn{1}{c|}{15.92} & 18.64 & 64.66 & 5.44 & \multicolumn{1}{c|}{11.65} & 38.83 & \multicolumn{1}{c|}{67.57} & 49.32 & \multicolumn{1}{c|}{85.44} & 60.19 & 90.29 & 0.97 & \multicolumn{1}{c|}{4.47} & 3.50 & 31.46 \\ \midrule

\tool & \textbf{7.57} & \multicolumn{1}{c|}{70.10} & \textbf{40.97} & \textbf{76.89} & \textbf{14.76} & \multicolumn{1}{c|}{\textbf{50.87}} & \textbf{51.84} & \multicolumn{1}{c|}{\textbf{83.88}} & \textbf{63.88} & \multicolumn{1}{c|}{\textbf{93.59}} & \textbf{69.90} & \textbf{95.92} & \textbf{17.67} & \multicolumn{1}{c|}{\textbf{50.29}} & \textbf{23.30} & \textbf{63.30} \\ \bottomrule
\end{tabular}
\vspace{-1em}
\end{table*}

\subsection{RQ1: Effectiveness of \tool}
\begin{figure}[htbp]
    \centering
    \includegraphics[width=0.43\textwidth]{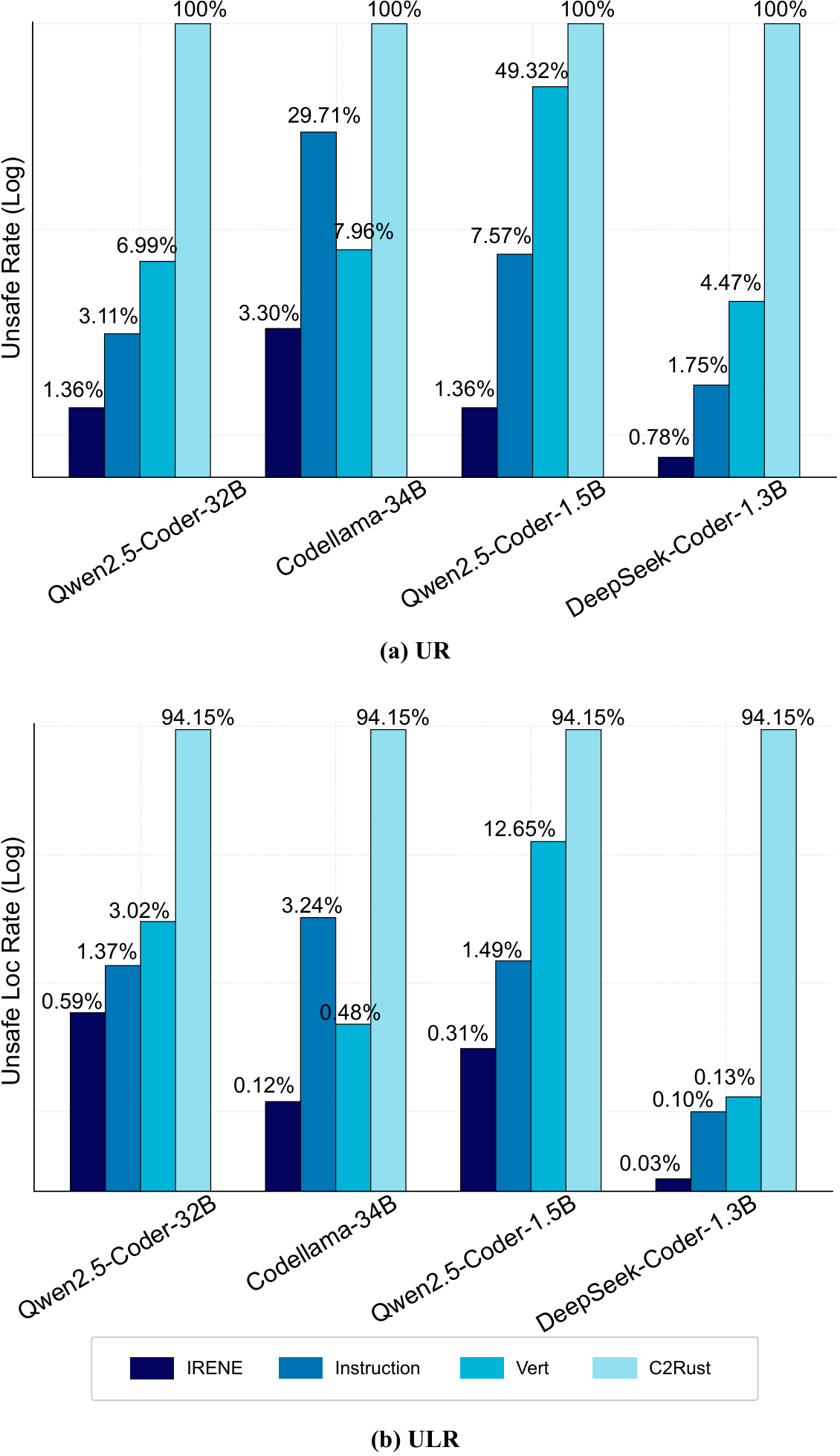}
    \caption{Comparison between \tool and three baselines on Unsafe Rate (UR) and Unsafe Loc Rate (ULR).}
    \label{Unsafe}
    \vspace{-1em}
\end{figure}

 % Comparison with Baselines
To investigate the effectiveness of \tool in C-to-Rust translation, we compare \tool with five representative prompt baselines on eight popular open-source LLMs. Table~\ref{effectiveness} presents the experimental results of \tool along with baseline methods. For the safety performance, we compare \tool with two static tools. Fig.~\ref{Unsafe} shows the safety performance.

\textbf{Comparison of the translation accuracy metrics.} 
As shown in Table~\ref{effectiveness}, \tool consistently outperforms all baseline methods across different LLMs. For example, compared to the strongest baseline method RAG, \tool achieves an average improvement of 8.06\% and 12.74\% on the CA and CSR metrics, respectively. These results demonstrate that \tool is more effective in generating correct and compilable Rust code. Moreover, when analyzing the improvements across different LLMs, we observe that \tool exhibits strong generalizability across models with varying parameter sizes. In particular, the gains are more substantial for larger models. 
\begin{figure}[htbp]
    \centering
    \begin{subfigure}[b]{0.23\textwidth}
        \centering
        \includegraphics[width=\textwidth]{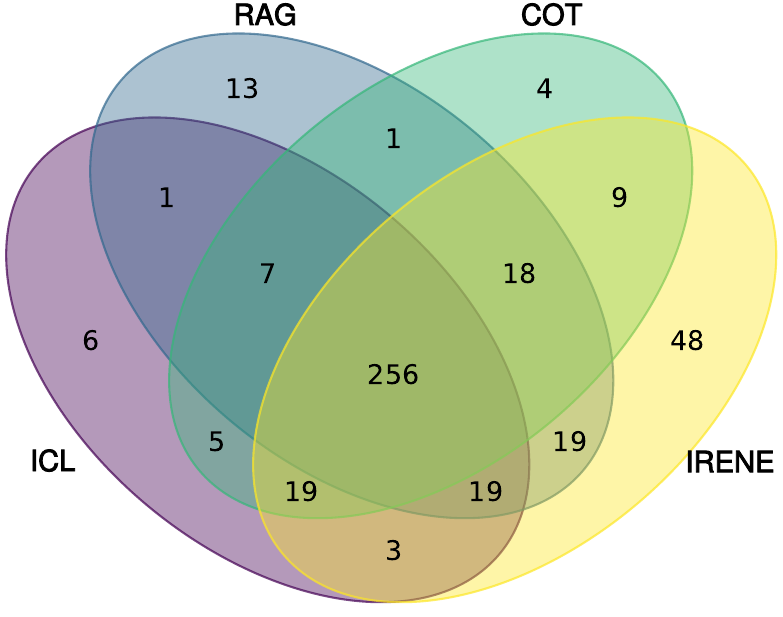}
        \caption{Qwen2.5-Coder-32B}
        \label{fig:sub1}
    \end{subfigure}
    \hspace{0.2cm}
    \begin{subfigure}[b]{0.23\textwidth}
        \centering
        \includegraphics[width=\textwidth]{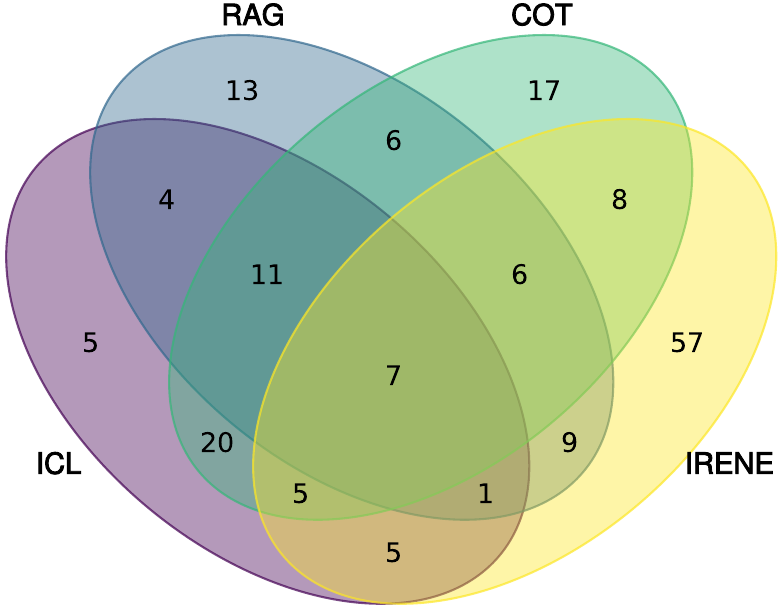}
        \caption{Qwen2.5-Coder-1.5B}
        \label{fig:sub2}
    \end{subfigure}
    \caption{Venn diagrams of successfully translated code generated by \tool without refinement and baseline methods on two different parameter sizes of LLMs.}
    \label{Venn}
\end{figure}
For instance, \tool improves CA by 2.48\% and 6.53\% on average for Deepseek-Coder-1.3B and Qwen2.5-Coder-1.5B, respectively, and achieves larger gains of 16.08\% and 15.73\% for Deepseek-Coder-6.7B and Qwen2.5-Coder-7B. This discrepancy can be explained by the limited instruction-following capability of smaller models, which often struggle to fully comprehend the rule hints and semantic guidance provided by \tool. 
% In contrast, larger models, such as XXX, are better equipped to understand and utilize the structured prompting, leading to more accurate and robust translations.

To quantify the accuracy of each translation method, we adopt an evaluation strategy commonly used in other works[], measuring the overlap of the correctly translated code between different methods. Figure~\ref{Venn} presents the Venn diagrams of successfully translated code generated by \tool without feedback and baseline methods. Regardless of parameter size, \tool consistently produces a larger number of uniquely correct translations. 
Notably, \tool identifies 48 and 57 unique correct translations in Qwen2.5-Coder-32B and Qwen2.5-Coder-1.5B, respectively, while other methods only range from 4 to 13 and 5 to 17. This not only demonstrates the generalizability of our framework, but also highlights that the rule-based guidance provided by \tool cannot be easily replicated by other prompting strategies.

\begin{table*}[t]
% \centering
\caption{Results of ablation study in translation accuracy. Under each metric, the best performance is highlighted in bold.}
\label{ablation}
\resizebox{\linewidth}{!}{%
\begin{tabular}{@{}l|cccc|cccccccc|cccc@{}}
\toprule
      & \multicolumn{4}{c|}{Deepseek-Coder} & \multicolumn{8}{c|}{Qwen2.5-Coder} & \multicolumn{4}{c}{Codellama} \\
 \cmidrule(lr){2-5} \cmidrule(lr){6-13}  \cmidrule(lr){14-17}
\multicolumn{1}{c|}{\textbf{Approach}} & \multicolumn{2}{c|}{1.3B} & \multicolumn{2}{c|}{6.7B} & \multicolumn{2}{c|}{1.5B} & \multicolumn{2}{c|}{7B} & \multicolumn{2}{c|}{14B} & \multicolumn{2}{c|}{32B} & \multicolumn{2}{c|}{7B} & \multicolumn{2}{c}{34B}\\
& CA & \multicolumn{1}{c|}{CSR} & CA & \multicolumn{1}{c|}{CSR} & CA & \multicolumn{1}{c|}{CSR} & CA & \multicolumn{1}{c|}{CSR} & CA & \multicolumn{1}{c|}{CSR} & CA & \multicolumn{1}{c|}{CSR} & CA & \multicolumn{1}{c|}{CSR} & CA & CSR  \\ \midrule

-w/o rule-augmented & 6.99 & \multicolumn{1}{l|}{33.79} & \multicolumn{1}{c}{24.85} & \multicolumn{1}{c|}{38.25} & 7.57 & \multicolumn{1}{l|}{15.34} & \multicolumn{1}{c}{46.41} & \multicolumn{1}{c|}{60.00} & \multicolumn{1}{c}{55.73} & \multicolumn{1}{c|}{67.18} & \multicolumn{1}{c}{65.83} & 72.62 & \multicolumn{1}{c}{6.41} & \multicolumn{1}{c|}{20.39} & \multicolumn{1}{c}{9.32} & 38.64 \\

-w/o summarization & 5.63 & \multicolumn{1}{l|}{59.61} & \multicolumn{1}{c}{40.58} & \multicolumn{1}{c|}{\textbf{78.64}} & \multicolumn{1}{c}{\textbf{15.34}} & \multicolumn{1}{c|}{46.21} & \multicolumn{1}{c}{\textbf{53.20}} & \multicolumn{1}{c|}{81.94} & \multicolumn{1}{c}{60.39} & \multicolumn{1}{c|}{89.90} & \multicolumn{1}{c}{69.13} & \multicolumn{1}{c}{95.15} & \multicolumn{1}{c}{16.70} & \multicolumn{1}{c|}{42.14} & \multicolumn{1}{c}{21.75} & \textbf{67.38} \\

-w/o feedback & 6.60 & \multicolumn{1}{l|}{58.45} & \multicolumn{1}{c}{34.37} & \multicolumn{1}{c|}{63.69} & \multicolumn{1}{c}{12.82} & \multicolumn{1}{c|}{39.61} & \multicolumn{1}{c}{44.27} & \multicolumn{1}{c|}{70.49} & \multicolumn{1}{c}{59.22} & \multicolumn{1}{c|}{81.75} & \multicolumn{1}{c}{66.02} & \multicolumn{1}{c}{87.77} & \multicolumn{1}{c}{15.15} & \multicolumn{1}{c|}{40.58} & \multicolumn{1}{c}{19.22} & 44.66 \\ \midrule

\tool & \textbf{7.57} & \multicolumn{1}{l|}{\textbf{70.10}} & \multicolumn{1}{c}{\textbf{40.97}} & \multicolumn{1}{c|}{76.89} & 14.76 & \multicolumn{1}{c|}{\textbf{50.87}} & 51.84 & \multicolumn{1}{c|}{\textbf{83.88}} & \textbf{63.88} & \multicolumn{1}{c|}{\textbf{93.59}} & \textbf{69.90} & \textbf{95.92} & \textbf{17.67} & \multicolumn{1}{c|}{\textbf{50.29}} & \textbf{23.30} & 63.30 \\ \bottomrule
\end{tabular}
}
\end{table*}
\begin{table*}[t]
\centering
\caption{An additional experiment about the summary's quality. "-better summary" indicates that each model’s summary is generated by Qwen2.5-Coder-32B. }
\label{summary}
\resizebox{\linewidth}{!}{%
\begin{tabular}{@{}l|cccc|cccccccc|cccc@{}}
\toprule
      & \multicolumn{4}{c|}{Deepseek-Coder} & \multicolumn{8}{c|}{Qwen2.5-Coder} & \multicolumn{4}{c}{Codellama} \\
 \cmidrule(lr){2-5} \cmidrule(lr){6-13}  \cmidrule(lr){14-17}
\multicolumn{1}{c|}{\textbf{Approach}} & \multicolumn{2}{c|}{1.3B} & \multicolumn{2}{c|}{6.7B} & \multicolumn{2}{c|}{1.5B} & \multicolumn{2}{c|}{7B} & \multicolumn{2}{c|}{14B} & \multicolumn{2}{c|}{32B} & \multicolumn{2}{c|}{7B} & \multicolumn{2}{c}{34B}\\
& CA & \multicolumn{1}{c|}{CSR} & CA & \multicolumn{1}{c|}{CSR} & CA & \multicolumn{1}{c|}{CSR} & CA & \multicolumn{1}{c|}{CSR} & CA & \multicolumn{1}{c|}{CSR} & CA & \multicolumn{1}{c|}{CSR} & CA & \multicolumn{1}{c|}{CSR} & CA & CSR  \\ \midrule

-w/o summarization & 5.63 & \multicolumn{1}{l|}{59.61} & \multicolumn{1}{c}{40.58} & \multicolumn{1}{c|} {\textbf{78.64}} & \multicolumn{1}{c}{15.34} & \multicolumn{1}{c|}{46.21} & \multicolumn{1}{c}{53.20} & \multicolumn{1}{c|}{81.94} & \multicolumn{1}{c}{60.39} & \multicolumn{1}{c|}{89.90} & \multicolumn{1}{c}{69.13} & \multicolumn{1}{c|}{95.15} & \multicolumn{1}{c}{16.70} & \multicolumn{1}{c|}{42.14} & \multicolumn{1}{c}{21.75} & \textbf{67.38} \\

-better summary & \multicolumn{1}{c}{\textbf{9.13}} & \multicolumn{1}{c|}{62.91} & \multicolumn{1}{c}{\textbf{41.55}} & \multicolumn{1}{c|}{74.56} & \multicolumn{1}{c}{\textbf{18.45}} & \multicolumn{1}{c|}{\textbf{52.43}} & \multicolumn{1}{c}{\textbf{54.56}} & \multicolumn{1}{c|}{\textbf{85.05}} & \multicolumn{1}{c}{\textbf{64.27}} & \multicolumn{1}{c|}{93.01} & \textbf{69.90} & \textbf{95.92} & 17.28 & \multicolumn{1}{c|}{45.83} & \textbf{25.83} & 61.94 \\  \midrule

\tool & 7.57 & \multicolumn{1}{l|}{\textbf{70.10}} & \multicolumn{1}{c}{40.97} & \multicolumn{1}{c|}{76.89} & 14.76 & \multicolumn{1}{c|}{50.87} & 51.84 & \multicolumn{1}{c|}{83.88} & 63.88 & \multicolumn{1}{c|}{\textbf{93.59}} & 69.90 & 95.92  & \textbf{17.67} & \multicolumn{1}{c|}{\textbf{50.29}} & 23.30 & 63.30 \\ \bottomrule
\end{tabular}
}
\end{table*}
\textbf{Comparison of the translation safety metrics.}
As shown in Fig.~\ref{Unsafe}, \tool significantly outperforms both traditional static and LLM-based baselines in terms of safety. For instance, \tool and Vert achieve an average reduction of 82.82\% and 98.30\% on the UR metric when compared with C2Rust. These results demonstrate the superiority of LLMs in generating safe Rust code. Moreover, when compared to Vert, another LLM-based framework, our method achieves an average reduction of 15.73\% on the UR. This indicates that even under the same LLM-based setting, \tool is more effective at producing safe Rust code. In addition, it consistently demonstrates this capability across models of different sizes. For example, on the larger Qwen2.5-Coder-32B model, \tool reduces the Unsafe Rate by 5.63\% compared to Vert, and on the smaller Qwen2.5-Coder-1.5B model, it achieves a reduction of 47.96\%.

In addition to the UR metric, we also evaluate the ULR, which reflects the proportion of unsafe tokens in the translated code. 
Similarly, \tool consistently achieves the best results compared to both C2Rust and Vert across different LLMs. For example, it yields an average reduction of 4.33\% compared to Vert. \tool also demonstrates strong performance across models of varying sizes---for instance, achieving a 2.43\% reduction on Qwen2.5-Coder-32B and 12.34\% on Qwen2.5-Coder-1.5B.
These results consistently demonstrate that \tool produces safer Rust code by effectively leveraging rule-based guidance, achieving both accuracy and safety in the translation process.
\begin{tcolorbox}
\textbf{Answer to RQ1:} 
\tool achieves notable improvements over baselines in both translation accuracy and safety across different LLMs. It improves CA and CSR by 8.06\% and 12.74\% over the strongest baseline, demonstrating superior translation accuracy. In terms of translation safety, it significantly outperforms both traditional static tools and LLM-based approaches---with an average UR and ULR reduced to just 1.70\% and 0.26\% on average, respectively.
\end{tcolorbox}

\subsection{RQ2: Ablation Study}
To validate the effectiveness of each module in our framework, we perform ablation experiments, as shown in Table~\ref{ablation}.

\textbf{Rule-Augmented Retrieval Module.} In this experiment, we remove the rule hints and examples used by LLM to assess the effectiveness of the rule-augmented retrieval module. As shown in Table~\ref{ablation}, this removal leads to a significant drop in both CA and CSR across all model sizes. For example, on Qwen2.5-Coder-1.5B, the CA drops from 14.76\% to 7.57\%, and the CSR drops from 50.87\% to 15.34\%. Similarly, on Qwen2.5-Coder-14B, CA decreases from 63.88\% to 55.73\%, and CSR from 93.59\% to 67.18\%. These results indicate that the rule-augmented module plays a crucial role in enhancing both the correctness and compilability of translated code. This suggests that the rule-augmented retrieval module helps LLMs better adapt to Rust’s strict rules.

\textbf{Structured Summarization Module.} To assess the effectiveness of the structured summarization module, we first remove the summary generated by this component. Experimental results show that the absence of the summary leads to performance degradation in most models. For example, in Qwen2.5-Coder-14B, the CA drops by 3.47\% and the CSR by 3.69\%. This highlights the importance of providing the LLM with its own understanding of the source code.

However, as shown in Table~\ref{ablation}, a few smaller models, such as Qwen2.5-Coder-1.5B and 7B, show a slight improvement in CA after removing the summary. To further investigate this counterintuitive result, we conducted an additional experiment in which we replaced each model’s self-generated summary with the one produced by a more powerful model (Qwen2.5-Coder-32B). As shown in Table~\ref{summary}, all models experience a performance drop when the summary is removed. For instance, it suffers a 3.11\% reduction in CA on Qwen2.5-Coder-1.5B. These results indicate that the effectiveness of the summary largely depends on its quality.

In conclusion, 
% the summary effectively enhances the LLM’s understanding of code semantics. Even for smaller models, the summary module remains broadly applicable. 
only in rare cases may the model's limited semantic understanding and generation capability prevent it from fully leveraging the benefits of the summary. In such scenarios, providing a higher-quality summary generated by a more powerful model can effectively compensate for this limitation.

\textbf{Error-driven Translation Module.}
\begin{figure*}[htbp]
    \vspace{-1em}
    \centering
    \includegraphics[scale=0.55]{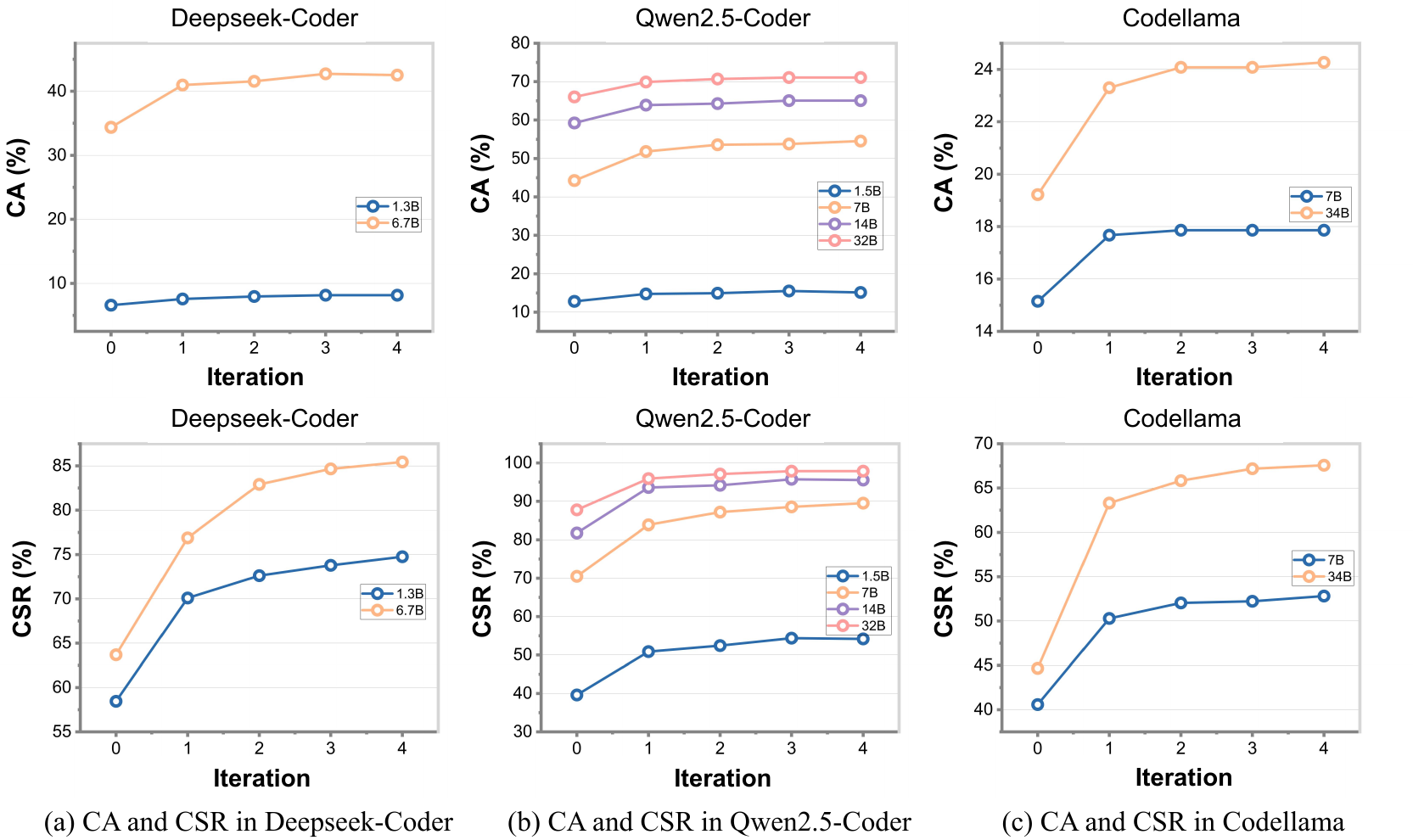}
    \caption{The result of iteration counts in CA and CSR metrics with different LLMs. }
    \label{iteration}
    \vspace{-1em}
\end{figure*}
This experiment is to reveal the impact of the error-driven translation module by removing the refinement. As shown in Table~\ref{ablation}, the absence of compilation error feedback leads to a noticeable decline in performance. For example, it suffers an average reduction of 4.08\% and 12.06\% in CA and CSR, respectively. These results highlight the strong self-correction capabilities of LLMs when provided with concrete error information, emphasizing the importance of iterative refinement during translation.

To further evaluate the effect of iteration count, we conduct additional experiments, with results shown in Fig.~\ref{iteration}. The results yield two main findings. First, increasing the number of iterations leads to consistent improvements in both CA and CSR. Second, the gains begin to plateau after three iterations, indicating diminishing returns from further refinement.
\begin{tcolorbox}
\textbf{Answer to RQ2:} 
All components in \tool contribute to the performance. Removing the rule-augmented retrieval module, structured summarization module, or error-driven translation module leads to substantial performance decreases.
\end{tcolorbox}

\subsection{RQ3: Performance in Industrial Scenarios}
\begin{table}[]
\centering
\caption{Comparison between \tool and baselines on CSR, UR and ULR in the \textbf{HW-Bench}. Under each metric, the best performance is highlighted in bold.}
\label{industry}
\resizebox{\linewidth}{!}{%
\begin{tabular}{@{}l|ccccccccc@{}}
\toprule
     & \multicolumn{9}{c}{Qwen2.5-Coder} \\
 \cmidrule(lr){2-10}
\multicolumn{1}{c|}{\textbf{Approach}} & \multicolumn{3}{c|}{7B} & \multicolumn{3}{c|}{14B} & \multicolumn{3}{c}{32B} \\
& CSR & UR & \multicolumn{1}{c|}{ULR} & CSR & UR & \multicolumn{1}{c|}{ULR} & CSR & UR & \multicolumn{1}{c}{ULR} \\ \midrule
Instruction & 20.00 & 25.00 & \multicolumn{1}{c|}{9.14} & 20.00 & 29.00 & \multicolumn{1}{c|}{9.83} & 27.00 & 33.00 & 15.17 \\

ICL & 26.00 & 35.00 & \multicolumn{1}{c|}{10.02} & 37.00 & 26.00 & \multicolumn{1}{c|}{9.18} & 32.00 & 34.00 & 14.33 \\

RAG & 36.00 & 38.00 & \multicolumn{1}{c|}{10.53} & 51.00 & 33.00 & \multicolumn{1}{c|}{11.60} & 41.00 & 36.00 & 14.04 \\

COT & \textbf{37.00} & 40.00 & \multicolumn{1}{c|}{11.00} & 46.00 & 28.00 & \multicolumn{1}{c|}{9.63} & 43.00 & 36.00 & 13.36 \\

Vert & 35.00 & 45.00 & \multicolumn{1}{c|}{14.59} & 53.00 & 43.00 & \multicolumn{1}{c|}{14.00} & 61.00 & 53.00 & 19.46 \\ \midrule

\tool & 30.00 & \textbf{23.00} & \multicolumn{1}{c|}{\textbf{7.87}} & \textbf{57.00} & \textbf{25.00} & \multicolumn{1}{c|}{\textbf{5.33}} & \textbf{63.00} & \textbf{15.00} & \textbf{3.37}  \\ \bottomrule
\end{tabular}
}
\vspace{-1em}
\end{table}
% Evaluation in Industrial Scenarios
To assess the applicability of \tool in the industrial scenario, we collaborate with Huawei and randomly select 100 C functions from their product-line codebase.
% , covering a broad range of business domains and functional modules. 
We apply \tool to translate these functions into Rust, and compare the results against several prompt-based baselines. Due to the limitations of time and resources, we conduct our experiment only on three open-source LLMs. The experimental result is shown in Table~\ref{industry}.

\textbf{Comparison of the translation accuracy metrics.} 
In the absence of practical verification methods, translation accuracy is assessed solely based on the CSR metric, with CA omitted. As shown in Table~\ref{industry}, \tool consistently achieves the highest CSR across three LLMs. For instance, it yields an average improvement of 22.20\% under Qwen2.5-Coder-32B, indicating strong robustness in handling diverse and complex industrial code. Although other methods may occasionally perform better in CSR, such as in Qwen2.5-Coder-7B, \tool still maintains the lowest UR and ULR.
% Notably, on Qwen2.5-Coder-32B, \tool achieves both the highest CSR and the best safety metrics, demonstrating its ability to balance correctness and safety effectively.
\begin{figure}[t]
    \centering
    \includegraphics[width=0.5\textwidth]{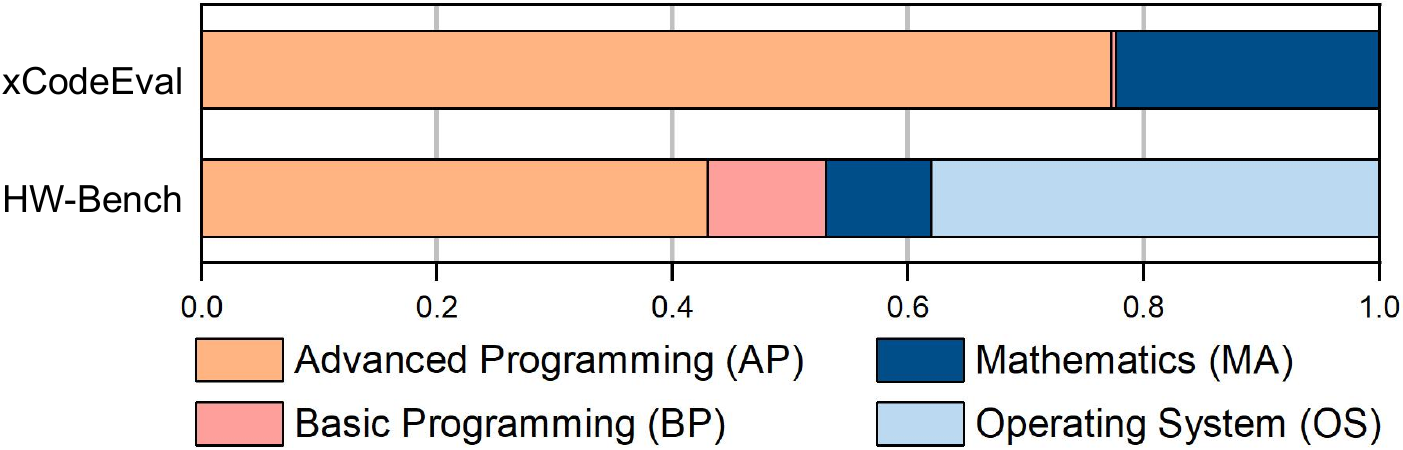}
    \caption{Application domain distributions of xCodeEval and HW-Bench.}
    \label{domain}
    \vspace{-2em}
\end{figure}

\textbf{Comparison of the translation safety metrics.} 
In terms of safety, Table~\ref{industry} demonstrates that \tool achieves the lowest UR and ULR across all LLMs, with an average reduction of 14.60\% and 6.87\%, respectively. These improvements highlight \tool's superiority in generating safer Rust code compared to five baselines.

Overall, the results indicate that \tool not only performs reliably in the public benchmark, but also generalizes well to industrial scenario. Moreover, following prior work~\cite{FullStack}, we analyze the application domain distributions of both xCodeEval and HW-Bench, as shown in Fig.~\ref{domain}. HW-Bench contains a higher proportion of operating system–related code, which involves system interactions. Despite these challenges, \tool maintains a strong safety performance.
\begin{tcolorbox}
\textbf{Answer to RQ3:} 
\tool shows the effectiveness in industrial scenarios. On HW-Bench, \tool achieves an average improvement of 12.33\% in CSR, compared with all baselines under employed LLMs. In terms of safety, it achieves an average reduction of 14.60\% in UR and 6.87\% in ULR compared to all baselines across all LLMs. These results demonstrate strong robustness and applicability in industrial codebases.
\end{tcolorbox}

\section{Discussion}
% \begin{figure*}
%     \centering
%     \includegraphics[width=1\textwidth]{Figures/case1.pdf}
%     \caption{Overview of \tool. In the Rule-Augmented Retrieval module, we }
%     \label{case1}
% \end{figure*}

% \begin{figure*}
%     \centering
%     \includegraphics[width=1\textwidth]{Figures/case2.pdf}
%     \caption{Overview of \tool. In the Rule-Augmented Retrieval module, we }
%     \label{case2}
% \end{figure*}

\subsection{Why does \tool work?}
\begin{figure}[t]
    \centering
    \includegraphics[width=0.45\textwidth]{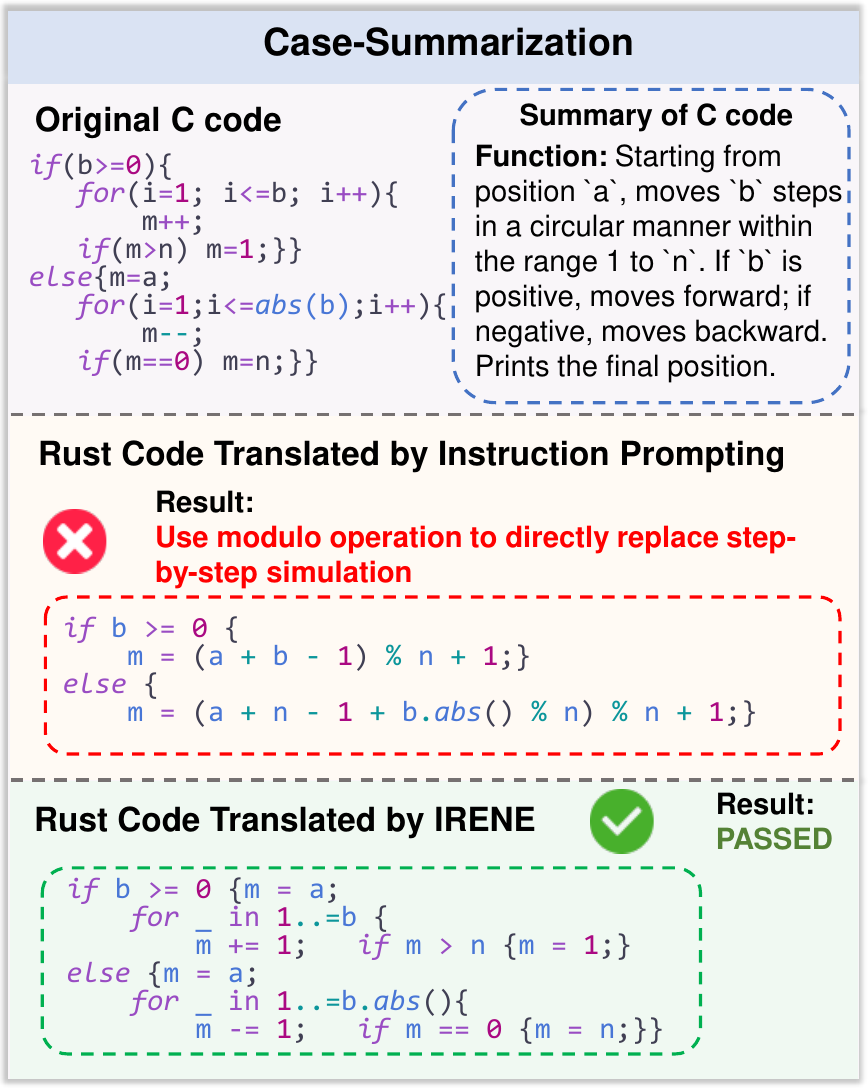}
    \caption{A case shows that how a structured summary ensures that the subsequent Rust translation remains semantically faithful. In this case, the LLM used is Qwen2.5-Coder-32B.}
    \label{case_summary}
\end{figure}

\begin{figure*}
    \centering
    \includegraphics[width=0.9\textwidth]{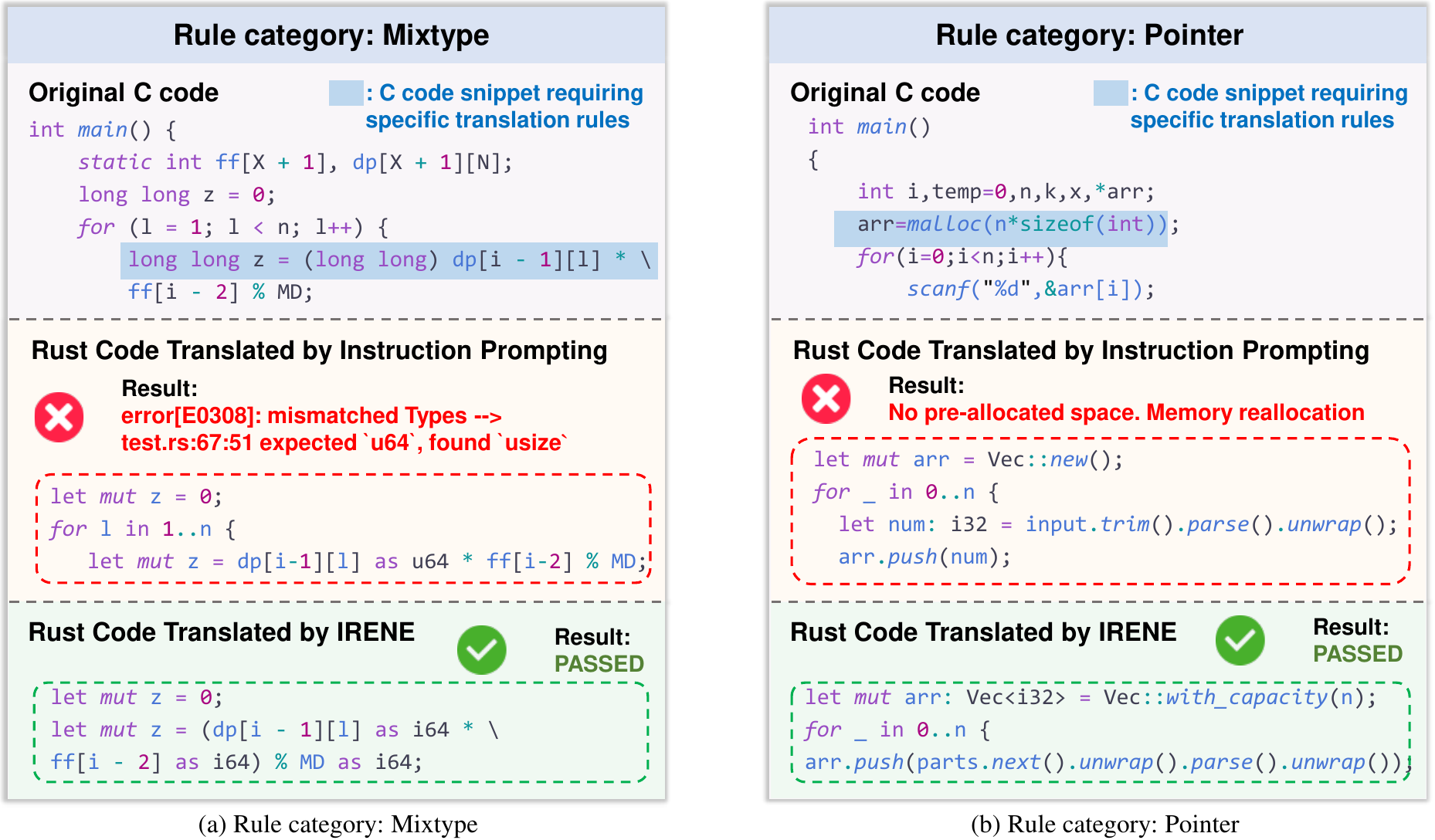}
    \caption{Two additional cases illustrate that how \tool improves translation results by addressing two different categories of rules based on Qwen2.5-Coder-32B.
% (a) In the Mixtype case, instruction prompting fails to resolve type mismatches caused by mixed integer types in Rust, leading to a compilation error. In contrast, \tool applies explicit type casting rules to ensure correct arithmetic operations across usize and i64.
% (b) In the Pointers case, the baseline translation allocates an empty Vec without pre-sizing, resulting in repeated memory reallocations during push. \tool detects the original C allocation semantics (malloc) and generates a Vec::with\_capacity accordingly, yielding more efficient and faithful translations.
}
    \label{rule_2}
\end{figure*}

In this section, we identify the following two advantages of \tool, which can explain its effectiveness and safety in C-to-Rust translation. For each advantage, we demonstrate the corresponding cases, as shown in Fig.~\ref{rule_1} and Fig.~\ref{case_summary}~-~\ref{rule_2}.
% , Figure~\ref{rule_2} and Figure~\ref{case_summary}. 

% \begin{figure*}
%     \centering
%     \begin{subfigure}[b]{0.47\textwidth}
%         \centering
%         \includegraphics[width=\textwidth]{Figures/case_IO.pdf}
%         \caption{Rule category: I/O}
%         \label{rule_io}
%     \end{subfigure}
%     \hspace{0.1cm}
%     \begin{subfigure}[b]{0.47\textwidth}
%         \centering
%         \includegraphics[width=\textwidth]{Figures/case_array.pdf}
%         \caption{Rule category: Array}
%         \label{rule_array}
%     \end{subfigure}
%     \caption{Two case showing how \tool improves translation quality by following Rust-specific rules.
% (a) In the I/O case, instruction prompting fails to handle multiple variable inputs correctly, resulting in a runtime panic. \tool adopts a rule-guided strategy using read\_to\_string and split\_whitespace, which ensures semantic equivalence and avoids the error.
% (b) In the Array case, the baseline produces a type mismatch error by indexing with i32. \tool resolves this by converting the index to usize, conforming to Rust's array indexing rule.}
%     \label{rule_1}
% \end{figure*}

\textbf{The ability to adapt to Rust's specific rules.}
The first advantage of \tool lies in its rule-based translation examples. LLMs heavily rely on pretraining knowledge.
% and 
Without explicit
% \yun{explicit}
% external 
guidance, it is quite difficult for them to accurately
% \yun{accurately}
% correctly 
apply Rust-specific rules during translation. \tool provides both rule hints and concrete translation examples to assist the model, guiding it to produce more accurate and idiomatic Rust code. As shown in Fig.\ref{rule_1} and Fig.\ref{rule_2}, the Rust code translated by \tool demonstrates better correctness and explicitly handles Rust-specific rules. Specifically, for different rule categories: \textbf{(1) I/O:} As shown in Fig.~\ref{rule_1} (a), the original C code uses \textit{scanf} to input multiple variables. The LLM under instruction prompting chooses \textit{read\_line}, which causes errors due to its semantic mismatch with \textit{scanf}. In contrast, \tool uses \textit{read\_to\_string} combined with \textit{split\_whitespace}, achieving a semantically equivalent conversion. \textbf{(2) Array:} Fig.~\ref{rule_1} (b) demonstrates that \tool correctly uses the \textit{usize} type for indexing arrays, rather than the incorrect \textit{i32} used by the LLM under instruction prompting. \textbf{(3) Mixtype:} As illustrated in Fig.~\ref{rule_2} (a), the C code uses explicit type casting (e.g., (long long)). The LLM under instruction prompting translation results in a type mismatch error (error[E0308]) due to inconsistent types within a single expression. In contrast, \tool successfully addresses this by applying appropriate type conversions. \textbf{(4) 
 Pointers:} As depicted in Fig.~\ref{rule_2} (b), the C code uses \textit{malloc} to allocate n \textit{int} memory for variable \textit{arr}. The LLM under instruction prompting simply uses \textit{Vec::new()}, while \tool enables the LLM to use \textit{Vec::with\_capacity(n)}, which better matches the original C's allocation semantics and avoids unnecessary reallocations.
% (The expression ``The LLM does..." specifically denotes instruction prompting.\yun{[???]})

% \begin{enumerate}

%     \item \textbf{I/O:} As shown in Fig.~\ref{rule_1} (a), the original C code uses \textit{scanf} to input multiple variables. The LLM under instruction prompting chooses \textit{read\_line}, which causes errors due to its semantic mismatch with \textit{scanf}. In contrast, \tool uses \textit{read\_to\_string} combined with \textit{split\_whitespace}, achieving a semantically equivalent conversion.

%     \item \textbf{Array:} Fig.~\ref{rule_1} (b) demonstrates that \tool correctly uses the \textit{usize} type for indexing arrays, rather than the incorrect \textit{i32} used by the LLM under instruction prompting.

%     \item \textbf{Mixtype:} As illustrated in Fig.~\ref{rule_2} (a), the C code uses explicit type casting (e.g., (long long)). The LLM under instruction prompting translation results in a type mismatch error (error[E0308]) due to inconsistent types within a single expression. In contrast, \tool successfully addresses this by applying appropriate type conversions.

%     \item \textbf{Pointers:} As depicted in Fig.~\ref{rule_2} (b), the C code uses \textit{malloc} to allocate n \textit{int} memory for variable \textit{arr}. The LLM under instruction prompting simply uses \textit{Vec::new()}, while \tool enables the LLM to use \textit{Vec::with\_capacity(n)}, which better matches the original C's allocation semantics and avoids unnecessary reallocations.

% \end{enumerate}

\textbf{The ability to better understand the code semantics.}
The second advantage of \tool lies in the code summary generated by the LLM itself.
Without the summary, the LLM focuses only on the surface structure of the code and tends to ignore critical semantic constraints. For instance, as illustrated in Fig~\ref{case_summary}, the model fails to recognize that the core logic involves the "circular manner" and instead attempts to simplify the loop using modulo operations, which leads to incorrect results.
In contrast, when a summary is provided, the clear task description enhances the semantic understanding of the code, guiding the model to generate more accurate translations. For example, in Fig~\ref{case_summary}, the summary explicitly identifies the ``circular manner'' behavior, effectively preventing the misuse of modulo operations.

\subsection{Threats to Validity}
For our study, we discuss three main threats to the validity:

\textbf{The selection of LLMs.} One potential threat to validity stems from the specific LLMs we evaluated. Since our primary goal is to enable deployment in industrial settings, we focused exclusively on eight popular open-source LLMs. This may limit the generalizability of our findings. In future work, we plan to extend our evaluation to include closed-source models such as ChatGPT and Gemini to further assess the general effectiveness of our framework.

\textbf{Potential data leakage}. Given that the pre-training data of Deepseek-Coder and Qwen2.5-Coder is not publicly disclosed, concerns about potential data leakage are understandable. However, we observe that when these models are prompted without our framework, the translation quality significantly degrades. This discrepancy strongly suggests that \tool’s performance gains are not attributable to memorized training content.

\textbf{Representativeness of datasets.} Another threat comes from the representativeness of the evaluation datasets. In this paper, we use two datasets: one is the open-source \textit{XCodeEval}, which primarily consists of programming competition-style code; the other is a manually collected set of functions from proprietary industrial projects. Due to confidentiality constraints, the industrial dataset cannot be released publicly, which may hinder reproducibility and broader evaluation. In future work, we plan to expand our evaluation to include more diverse and representative codebases to further validate the generality of \tool.

\section{Related Work}
\subsection{Rule-based C-to-Rust translation}
Rule-based translation approaches rely on predefined transformation rules and compiler tools to convert C code into structurally equivalent Rust code. These methods typically perform AST-level level transformations, mapping C constructs—including pointers, structs, and manual memory management—into Rust’s corresponding syntax and APIs.
For example, the widely used tool C2Rust~\cite{immunant2022c2rust}, built on Clang and LLVM, translates C code into \emph{unsafe} Rust through C-Rust FFI. Subsequent work~\cite{concrat,ownership,Donotwrite,Emre2021,Emre2023}, builds on the C2Rust tool, introducing improvements from various perspectives. Emre et al.~\cite{Emre2021,Emre2023} proposes the pseudo-safety technique, which addresses the problem of dereferencing raw pointers by emulating C-style aliasing behavior using Rust’s raw pointers. Concrat~\cite{concrat} developed a static analyzer specifically designed to handle Lock API. Crown~\cite{ownership} defines a set of ownership rules to guide the conversion of pointer types. In contrast to these tools, \tool combines predefined static analysis rules with summarization of the code to enhance the automation and precision of rewriting.

\subsection{LLM-based C-to-Rust translation}
LLM-based approaches employ LLMs to translate C programs into Rust by capturing patterns from programming corpora. 
LLM-based approaches incorporate structural and semantic information into prompts to guide translation, and apply feedback to refine outputs.
% These models are typically trained or fine-tuned on bilingual code corpora, allowing them to learn structural and semantic correspondences across programming languages. 
% 这段的方法都是做翻译的
For instance, Vert~\cite{VERT} uses LLMs to generate Rust code from C with a feedback-based strategy. Syzygy~\cite{Syzygy} combines LLMs with sampling and test-based feedback. C2SaferRust~\cite{C2S} integrates call graph and data flow analysis with LLMs and feedback repair. SACTOR~\cite{Multi} leverages static tools such as C2Rust and Crown to assist LLMs. Compared to these approaches, \tool enhances both static rule and semantic understanding, and produces more reliable and safer Rust code.
% 这段的方法则是专注某一方面的

In addition to these translation methods, some LLM-based studies focus on specific aspects. For example, Hong et al.\cite{signature} target the translation of function signatures. Eniser et al.\cite{crownllm} apply fuzz testing to validate translation correctness. Li et al.~\cite{user} conduct a user study to investigate how human experts perform code translation.

\section{Conclusion}
In this paper, we propose \tool, an LLM-based framework for C-to-Rust translation. \tool enhances the capabilities of large language models by integrating both rule-based guidance and semantic understanding. It consists of three key components: a rule-augmented retrieval module, a structured summarization module, and an error-driven translation module. Extensive experiments demonstrate the effectiveness of \tool across various models and settings. Ablation studies further highlight the unique contributions of each component. In future work, we plan to extend \tool to support more programming languages, such as C++, and to tackle larger-scale translation tasks at the project level.

\bibliographystyle{ieeetr}
\bibliography{Main}

\begin{thebibliography}{10}

\bibitem{DBLP:conf/icse/Hong23}
J.~Hong, ``Improving automatic c-to-rust translation with static analysis,'' in {\em {ICSE} Companion}, pp.~273--277, {IEEE}, 2023.

\bibitem{DBLP:conf/apsys/ChenMWZZK11}
H.~Chen, Y.~Mao, X.~Wang, D.~Zhou, N.~Zeldovich, and M.~F. Kaashoek, ``Linux kernel vulnerabilities: state-of-the-art defenses and open problems,'' in {\em APSys}, p.~5, {ACM}, 2011.

\bibitem{DBLP:conf/imc/DurumericKAHBLWABPP14}
Z.~Durumeric, J.~Kasten, D.~Adrian, J.~A. Halderman, M.~D. Bailey, F.~Li, N.~Weaver, J.~Amann, J.~Beekman, M.~Payer, and V.~Paxson, ``The matter of heartbleed,'' in {\em Internet Measurement Conference}, pp.~475--488, {ACM}, 2014.

\bibitem{herman2019rewriting}
D.~Herman, ``Rewriting a browser component in rust.'' \url{https://hacks.mozilla.org/2019/02/rewriting-a-browser-component-in-rust/}, 2019.
\newblock Accessed: 2025-05-19.

\bibitem{rustforlinux_nova}
{Rust for Linux}, ``Nova: a rust-based gpu driver.'' \url{https://rust-for-linux.com/nova-gpu-driver}, 2024.
\newblock Accessed: 2025-05-19.

\bibitem{google2021rust}
{Google Security Blog}, ``Rust in the android platform.'' \url{https://security.googleblog.com/2021/04/rust-in-android-platform.html}, 2021.
\newblock Accessed: 2025-05-19.

\bibitem{DBLP:conf/cav/ZhangDYW23}
H.~Zhang, C.~David, Y.~Yu, and M.~Wang, ``Ownership guided {C} to rust translation,'' in {\em {CAV} {(3)}}, vol.~13966 of {\em Lecture Notes in Computer Science}, pp.~459--482, Springer, 2023.

\bibitem{DBLP:conf/kbse/HongR24}
J.~Hong and S.~Ryu, ``To tag, or not to tag: Translating c's unions to rust's tagged unions,'' in {\em {ASE}}, pp.~40--52, {ACM}, 2024.

\bibitem{DBLP:conf/icse/LingYWWCH22}
M.~Ling, Y.~Yu, H.~Wu, Y.~Wang, J.~R. Cordy, and A.~E. Hassan, ``In rust we trust - {A} transpiler from unsafe {C} to safer rust,'' in {\em ICSE-Companion}, pp.~354--355, {ACM/IEEE}, 2022.

\bibitem{immunant2022c2rust}
Immunant, ``C2rust.'' \url{https://github.com/immunant/c2rust}, 2022.

\bibitem{clang}
{LLVM Project}, ``Clang: A c language family frontend for llvm.'' \url{https://clang.llvm.org/}.
\newblock Accessed: 2025-05-26.

\bibitem{DBLP:conf/cgo/LattnerA04}
C.~Lattner and V.~S. Adve, ``{LLVM:} {A} compilation framework for lifelong program analysis {\&} transformation,'' in {\em 2nd {IEEE} / {ACM} International Symposium on Code Generation and Optimization {(CGO} 2004), 20-24 March 2004, San Jose, CA, {USA}}, pp.~75--88, {IEEE} Computer Society, 2004.

\bibitem{concrat}
J.~Hong and S.~Ryu, ``Concrat: An automatic c-to-rust lock {API} translator for concurrent programs,'' in {\em 45th {IEEE/ACM} International Conference on Software Engineering, {ICSE} 2023, Melbourne, Australia, May 14-20, 2023}, pp.~716--728, {IEEE}, 2023.

\bibitem{ownership}
H.~Zhang, C.~David, Y.~Yu, and M.~Wang, ``Ownership guided {C} to rust translation,'' in {\em Computer Aided Verification - 35th International Conference, {CAV} 2023, Paris, France, July 17-22, 2023, Proceedings, Part {III}} (C.~Enea and A.~Lal, eds.), vol.~13966 of {\em Lecture Notes in Computer Science}, pp.~459--482, Springer, 2023.

\bibitem{Emre2021}
M.~Emre, R.~Schroeder, K.~Dewey, and B.~Hardekopf, ``Translating {C} to safer rust,'' {\em Proc. {ACM} Program. Lang.}, vol.~5, no.~{OOPSLA}, pp.~1--29, 2021.

\bibitem{Emre2023}
M.~Emre, P.~Boyland, A.~Parekh, R.~Schroeder, K.~Dewey, and B.~Hardekopf, ``Aliasing limits on translating {C} to safe rust,'' {\em Proc. {ACM} Program. Lang.}, vol.~7, no.~{OOPSLA1}, pp.~551--579, 2023.

\bibitem{Hong_static}
J.~Hong, ``Improving automatic c-to-rust translation with static analysis,'' in {\em 45th {IEEE/ACM} International Conference on Software Engineering: {ICSE} 2023 Companion Proceedings, Melbourne, Australia, May 14-20, 2023}, pp.~273--277, {IEEE}, 2023.

\bibitem{DBLP:journals/corr/abs-2409-10506}
M.~Shiraishi and T.~Shinagawa, ``Context-aware code segmentation for c-to-rust translation using large language models,'' {\em CoRR}, vol.~abs/2409.10506, 2024.

\bibitem{DBLP:journals/corr/abs-2412-14234}
M.~Shetty, N.~Jain, A.~Godbole, S.~A. Seshia, and K.~Sen, ``Syzygy: Dual code-test {C} to (safe) rust translation using llms and dynamic analysis,'' {\em CoRR}, vol.~abs/2412.14234, 2024.

\bibitem{DBLP:journals/corr/abs-2501-14257}
V.~Nitin, R.~Krishna, L.~L. do~Valle, and B.~Ray, ``C2saferrust: Transforming {C} projects into safer rust with neurosymbolic techniques,'' {\em CoRR}, vol.~abs/2501.14257, 2025.

\bibitem{VERT}
A.~Z.~H. Yang, Y.~Takashima, B.~Paulsen, J.~Dodds, and D.~Kroening, ``Vert: Verified equivalent rust transpilation with large language models as few-shot learners,'' 2024.

\bibitem{crownllm}
H.~F. Eniser, H.~Zhang, C.~David, M.~Wang, M.~Christakis, B.~Paulsen, J.~Dodds, and D.~Kroening, ``Towards translating real-world code with llms: {A} study of translating to rust,'' {\em CoRR}, vol.~abs/2405.11514, 2024.

\bibitem{C2S}
V.~Nitin, R.~Krishna, L.~L. do~Valle, and B.~Ray, ``C2saferrust: Transforming {C} projects into safer rust with neurosymbolic techniques,'' {\em CoRR}, vol.~abs/2501.14257, 2025.

\bibitem{Multi}
T.~Zhou, H.~Lin, S.~Jha, M.~Christodorescu, K.~Levchenko, and V.~Chandrasekaran, ``Llm-driven multi-step translation from {C} to rust using static analysis,'' {\em CoRR}, vol.~abs/2503.12511, 2025.

\bibitem{yang2024}
Z.~Yang, F.~Liu, Z.~Yu, J.~W. Keung, J.~Li, S.~Liu, Y.~Hong, X.~Ma, Z.~Jin, and G.~Li, ``Exploring and unleashing the power of large language models in automated code translation,'' {\em Proc. {ACM} Softw. Eng.}, vol.~1, no.~{FSE}, pp.~1585--1608, 2024.

\bibitem{xcode}
M.~A.~M. Khan, M.~S. Bari, X.~D. Long, W.~Wang, M.~R. Parvez, and S.~Joty, ``Xcodeeval: An execution-based large scale multilingual multitask benchmark for code understanding, generation, translation and retrieval,'' in {\em Proceedings of the 62nd Annual Meeting of the Association for Computational Linguistics (Volume 1: Long Papers), {ACL} 2024, Bangkok, Thailand, August 11-16, 2024} (L.~Ku, A.~Martins, and V.~Srikumar, eds.), pp.~6766--6805, Association for Computational Linguistics, 2024.

\bibitem{repo}
G.~Ou, M.~Liu, Y.~Chen, X.~Peng, and Z.~Zheng, ``Repository-level code translation benchmark targeting rust,'' {\em CoRR}, vol.~abs/2411.13990, 2024.

\bibitem{IRENE}
{IRENE}, ``Replication package of irene.'' \url{https://anonymous.4open.science/r/IRENE-2E0E}, 2025.

\bibitem{BM25}
S.~E. Robertson and H.~Zaragoza, ``The probabilistic relevance framework: {BM25} and beyond,'' {\em Found. Trends Inf. Retr.}, vol.~3, no.~4, pp.~333--389, 2009.

\bibitem{DBLP:conf/nips/BrownMRSKDNSSAA20}
T.~B. Brown, B.~Mann, N.~Ryder, M.~Subbiah, J.~Kaplan, P.~Dhariwal, A.~Neelakantan, P.~Shyam, G.~Sastry, A.~Askell, S.~Agarwal, A.~Herbert{-}Voss, G.~Krueger, T.~Henighan, R.~Child, A.~Ramesh, D.~M. Ziegler, J.~Wu, C.~Winter, C.~Hesse, M.~Chen, E.~Sigler, M.~Litwin, S.~Gray, B.~Chess, J.~Clark, C.~Berner, S.~McCandlish, A.~Radford, I.~Sutskever, and D.~Amodei, ``Language models are few-shot learners,'' in {\em NeurIPS}, 2020.

\bibitem{DBLP:conf/kdd/FanDNWLYCL24}
W.~Fan, Y.~Ding, L.~Ning, S.~Wang, H.~Li, D.~Yin, T.~Chua, and Q.~Li, ``A survey on {RAG} meeting llms: Towards retrieval-augmented large language models,'' in {\em {KDD}}, pp.~6491--6501, {ACM}, 2024.

\bibitem{DBLP:conf/nips/Wei0SBIXCLZ22}
J.~Wei, X.~Wang, D.~Schuurmans, M.~Bosma, B.~Ichter, F.~Xia, E.~H. Chi, Q.~V. Le, and D.~Zhou, ``Chain-of-thought prompting elicits reasoning in large language models,'' in {\em Advances in Neural Information Processing Systems 35: Annual Conference on Neural Information Processing Systems 2022, NeurIPS 2022, New Orleans, LA, USA, November 28 - December 9, 2022} (S.~Koyejo, S.~Mohamed, A.~Agarwal, D.~Belgrave, K.~Cho, and A.~Oh, eds.), 2022.

\bibitem{Deepseekcoder}
D.~Guo, Q.~Zhu, D.~Yang, Z.~Xie, K.~Dong, W.~Zhang, G.~Chen, X.~Bi, Y.~Wu, Y.~K. Li, F.~Luo, Y.~Xiong, and W.~Liang, ``Deepseek-coder: When the large language model meets programming - the rise of code intelligence,'' {\em CoRR}, vol.~abs/2401.14196, 2024.

\bibitem{hui2024qwen2}
B.~Hui, J.~Yang, Z.~Cui, J.~Yang, D.~Liu, L.~Zhang, T.~Liu, J.~Zhang, B.~Yu, K.~Dang, {\em et~al.}, ``Qwen2. 5-coder technical report,'' {\em arXiv preprint arXiv:2409.12186}, 2024.

\bibitem{Codallama}
B.~Rozi{\`{e}}re, J.~Gehring, F.~Gloeckle, S.~Sootla, I.~Gat, X.~E. Tan, Y.~Adi, J.~Liu, T.~Remez, J.~Rapin, A.~Kozhevnikov, I.~Evtimov, J.~Bitton, M.~Bhatt, C.~Canton{-}Ferrer, A.~Grattafiori, W.~Xiong, A.~D{\'{e}}fossez, J.~Copet, F.~Azhar, H.~Touvron, L.~Martin, N.~Usunier, T.~Scialom, and G.~Synnaeve, ``Code llama: Open foundation models for code,'' {\em CoRR}, vol.~abs/2308.12950, 2023.

\bibitem{huggingface}
{Hugging Face, Inc.}, ``Hugging face: The platform for machine learning and nlp.'' \url{https://huggingface.co/}.
\newblock Accessed: 2025-05-26.

\bibitem{pytorch}
{PyTorch Contributors}, ``Pytorch: An open source machine learning framework.'' \url{https://pytorch.org/}.
\newblock Accessed: 2025-05-26.

\bibitem{DBLP:conf/kbse/GaoWGWZL23}
S.~Gao, X.~Wen, C.~Gao, W.~Wang, H.~Zhang, and M.~R. Lyu, ``What makes good in-context demonstrations for code intelligence tasks with llms?,'' in {\em 38th {IEEE/ACM} International Conference on Automated Software Engineering, {ASE} 2023, Luxembourg, September 11-15, 2023}, pp.~761--773, {IEEE}, 2023.

\bibitem{threshold1}
P.~Bailey, R.~W. White, H.~Liu, and G.~Kumaran, ``Mining historic query trails to label long and rare search engine queries,'' {\em {ACM} Trans. Web}, vol.~4, no.~4, pp.~15:1--15:27, 2010.

\bibitem{FullStack}
Y.~Cheng, J.~Chen, J.~Chen, L.~Chen, L.~Chen, W.~Chen, Z.~Chen, S.~Geng, A.~Li, B.~Li, B.~Li, L.~Li, B.~Liu, J.~Liu, K.~Liu, Q.~Liu, S.~Liu, S.~Liu, T.~Liu, T.~Liu, Y.~Liu, R.~Long, J.~Mai, G.~Ning, Z.~Y. Peng, K.~Shen, J.~Su, J.~Su, T.~Sun, Y.~Sun, Y.~Tao, G.~Wang, S.~Wang, X.~Wang, Y.~Wang, Z.~Wang, J.~Xia, L.~Xiang, X.~Xiao, Y.~Xiao, C.~Xi, S.~Xin, J.~Xu, S.~Xu, H.~Yang, J.~Yang, Y.~Yang, J.~Yuan, J.~Zhang, Y.~Zhang, Y.~Zhang, S.~Zheng, H.~Zhu, and M.~Zhu, ``Fullstack bench: Evaluating llms as full stack coders,'' {\em CoRR}, vol.~abs/2412.00535, 2024.

\bibitem{Donotwrite}
J.~Hong and S.~Ryu, ``Don't write, but return: Replacing output parameters with algebraic data types in c-to-rust translation,'' {\em Proc. {ACM} Program. Lang.}, vol.~8, no.~{PLDI}, pp.~716--740, 2024.

\bibitem{Syzygy}
M.~Shetty, N.~Jain, A.~Godbole, S.~A. Seshia, and K.~Sen, ``Syzygy: Dual code-test {C} to (safe) rust translation using llms and dynamic analysis,'' {\em CoRR}, vol.~abs/2412.14234, 2024.

\bibitem{signature}
J.~Hong and S.~Ryu, ``Type-migrating c-to-rust translation using a large language model,'' {\em Empir. Softw. Eng.}, vol.~30, no.~1, p.~3, 2025.

\bibitem{user}
R.~Li, B.~Wang, T.~Li, P.~Saxena, and A.~Kundu, ``Translating {C} to rust: Lessons from a user study,'' in {\em 32nd Annual Network and Distributed System Security Symposium, {NDSS} 2025, San Diego, California, USA, February 24-28, 2025}, The Internet Society, 2025.

\end{thebibliography}

\end{document}